\documentclass[smallextended]{svjour3}

\usepackage{float}
\usepackage{hyperref}
\usepackage{mathptmx}       
\usepackage{helvet}         
\usepackage{courier}        
\usepackage{type1cm}        
\usepackage{makeidx}         

\usepackage{amscd,amssymb,amsmath,latexsym,bm}
\usepackage[mathcal,mathscr]{euscript}
\usepackage{lipsum}
\usepackage{amsfonts}
\usepackage{graphicx}
\usepackage{epstopdf}
\usepackage{algorithmic}
\usepackage{hyperref}
\usepackage{xcolor}
\usepackage{enumerate}
\usepackage{ulem}

\numberwithin{equation}{section}
\numberwithin{figure}{section}

\newcommand{\NM}{{\mathbb N}}
\newcommand{\RM}{{\mathbb R}}

\newcommand{\ZM}{{\mathbb Z}}
\newcommand{\PM}{{\mathbb P}}

\newcommand{\GM}{{\mathbb G}}

\newcommand{\Pp}{{\mathcal P}}

\newcommand{\Ss}{{\mathcal S}}
\newcommand{\Oo}{{\mathcal O}}

\newcommand{\Nn}{{\mathcal N}}

\newcommand{\Ll}{{\mathcal L}}

\newcommand{\bra}[1]{\langle #1|}
\newcommand{\ket}[1]{|#1\rangle}
\newcommand{\braket}[2]{\langle #1|#2\rangle}        

\begin{document}

\title{Disordered Crystals from First Principles I: Quantifying the Configuration Space}
\author{Thomas D. K\"uhne \and Emil Prodan}

\institute{Thomas D. K\"uhne \at
              Department of Chemistry, University of Paderborn, Paderborn, Germany \\
              \email{tdkuehne@mail.upb.de}            
           \and
           Emil Prodan \at
              Department of Physics, Yeshiva University, New York, New York, USA \\
              \email{prodan@yu.edu}
}

\maketitle

\begin{abstract}
This work represents the first chapter of a project on the foundations of first-principle calculations of the electron transport in crystals at finite temperatures. We are interested in the range of temperatures, where most electronic components operate, that is, room temperature and above. The aim is a predictive first-principle formalism that combines {\it ab-initio} molecular dynamics and a finite-temperature Kubo-formula for homogeneous thermodynamic phases. The input for this formula is the ergodic dynamical system $(\Omega,\GM,{\rm d}\PM)$ defining the thermodynamic crystalline phase, where $\Omega$ is the configuration space for the atomic degrees of freedom, $\GM$ is the space group acting on $\Omega$ and ${\rm d}\PM$ is the ergodic Gibbs measure relative to the $\GM$-action. The present work develops an algorithmic method for quantifying $(\Omega,\GM,{\rm d}\PM)$ from first principles. Using the silicon crystal as a working example, we find the Gibbs measure to be extremely well characterized by a multivariate normal distribution, which can be quantified using a small number of parameters. The latter are computed at various temperatures and communicated in the form of a table. Using this table, one can generate large and accurate thermally-disordered atomic configurations to serve, for example, as input for subsequent simulations of the electronic degrees of freedom.
\end{abstract}


\maketitle

\setcounter{tocdepth}{3}

\tableofcontents

\section{Introduction}

The investigation of the temperature dependence of transport coefficients in both metals and semiconductors has a long history \cite{FluggeI1956,FluggeII1957,Jacoboni1977,Acherman2011}. There is always a strong interest in refining the quantitative characterizations of the conductivity tensor of a material as a function of temperature \cite{Jacoboni1977,Acherman2011} and mapping the various temperature regimes,  where the transport coefficients have qualitatively distinct behaviors, is of great importance. For example, the surfaces of topological insulators are predicted to display ballistic charge transport even at room temperature and an immense experimental effort has been dedicated to prove this conjecture \cite{ORV}. Despite the intense interest from the academic and industrial communities, there are virtually no predictive theoretical methods to compute the transport coefficients of solids at finite temperatures from first principles. The present work represents the first step of a long term effort undertaken by the authors to fill this void.

\vspace{0.2cm}

The formalism we plan to adopt for our first-principle computational program is a combination of the quantum theory of transport for generic homogeneous condensed matter, perfected by Bellissard and collaborators \cite{BES,SBB1,SBB2}, and the second generation Car-Parrinello molecular dynamics (CPMD) method \cite{CPMD2,TDK2014}. The transport theory we just alluded to, culminates with a compact Kubo-formula for homogeneous phases at finite temperatures, which accounts for the dissipation too. The input for this formula consists of: I) The configuration space and the Gibbs measure for the atomic degrees of freedom; II) The family of Hamiltonians for the electronic degrees of freedom; III) The dissipation super-operator. Given these inputs, a controlled and exponentially fast (w.r.t. crystal's size) converging numerical implementation of the formalism has been developed in \cite{Pro1,Pro2} and, for simple models of disordered crystals, applications can be found in \cite{PB,SP,XP1,XP2}. The transport formalism and its numerical implementation has been extended in \cite{CancesJMP2017} to other aperiodic systems, such as incommensurately layered materials. Exploiting the superior efficiency of the novel second generation CPMD algorithm, which, for the system analyzed in the present work, results in a speed-up of two orders of magnitude, enabled the treatment of relatively large crystals and computations of rather long orbits. These new developments give us hope that, on one hand, the inputs I)-III) can be resolved from first principles and, on the other hand, that the Kubo-formula can be evaluated with sufficient precision.    

\vspace{0.2cm}

There is still much to learn about the transport characteristics of real crystals. Most electronic devices operate at room temperature and, even under normal loads, the actual operating temperatures can shoot up by tens or even hundreds of degrees. The existing microscopic theories of electron transport at finite temperatures build on the quantum electron states computed for the ideal zero-temperature periodic atomic configurations of the crystal \cite{Mah2000}. They also assume a weak coupling with the dissipative channels, such as the phonons, with the latter being computed in the linear waves regime, relative to the ideal zero-temperature atomic configurations. However, at room temperature and above, the ideal zero-temperature atomic configurations can not be used as a reference, for both electrons and phonons, because of significant effects that can change qualitatively the nature of the states. Indeed, in such conditions, the ionic cores undergo a vigorous thermal motions and the electrons have to navigate in a disordered environment. This thermally induced disorder can have fundamental effects on the electron transport. 

\vspace{0.2cm}

First, it can Anderson-localize the electron states near the energy band edges \cite{AAL}, and, for semiconductors, it can fill the insulating energy gap with localized impurity states. Even though the electron spectrum displays no gaps in this regime, in a dc-transport measurement, upon gating, one still observes activated behavior for the direct conductivity. This is because there are sharp energy levels that separate the extended electron states from the localized ones, the so called mobility edges. The latter define the mobility gap, which is the most important spectral parameter for transport. It is quite well documented that the mobility gap and the electron density of states can have a highly non-monotonic behavior with the strength of disorder \cite{LCJ,YNIK,Prodan2011,Monserrat2016}, and, for this reason, the dependence of the transport coefficients on temperature can be highly non-trivial and is hence difficult to account for, quantitatively, with phenomenological models. 

\vspace{0.2cm}

Secondly, because the ionic cores are quite far from the ideal equilibrium positions during the thermal motion, the phonons, which provide the main dissipation channels at room temperature and above, cannot be defined relative to the periodic configuration of the atoms and the use of zero-temperature phonon spectra and the density of states in electron transport simulations is questionable. The phonons need to be redefined and computed relative to the disordered atomic configurations and there is a large probability that these phonon modes are actually Anderson-localized, in which case they cannot diffuse and will be less effective at dissipating energy. This is again a significant effect that will be difficult to account for with phenomenological models.

\vspace{0.2cm}

All the above effects can be captured quantitatively by our proposed formalism. However, our present study is focused entirely on quantifying the configuration space and the Gibbs measure of the atomic degrees of freedom from first principles. We chose the pure silicon crystal to be our case study because it is the most important semiconductor for electronics industry and also because its transport coefficients display non-trivial behavior with temperature \cite{Jacoboni1977}. The main challenge for us was to devise an effective method to analyze the complex orbits in the highly multi-dimensional space of the atomic degrees of freedom. Our solution is to project these orbits on different planes of coordinates and in this way to generate a hierarchy of increasingly accurate representations of both the configuration space and Gibbs measure. One of our starting questions was: Can we encode the data in a small number of numerical parameters in order to efficiently communicate the results? The main finding of our work is the affirmative answer to this question. Quite remarkably, for all temperatures inside the range $300-1500\ K$, the Gibbs measure is found to be extremely well approximated by a normal multivariate distribution with correlations only up to $4^{\rm th}$-near neighbors. As a consequence, for a fixed temperature within the range mentioned above, the full Gibbs measure can be encoded in just five numerical parameters (see Table~\ref{Tab-Sigma}). These parameters appear to have a smooth dependence on the temperature (see Fig.~\ref{Fig-SigmaVsTemp}), hence the discrete set of temperatures examined in our study can be interpolated to a whole continuous range of temperatures. As such, our study provides a complete parametrization of the atomic configurations of the silicon crystal in the temperature range between $300\ K$ and $1500\ K$.

\vspace{0.2cm} 

The above conclusions are extremely important because, with such parameters at hand, anyone can generate meaningful and accurate thermally-disordered configurations of the crystal, without the need to repeat our extensive CPMD simulations. Furthermore, as we will demonstrate in our concluding remarks, these parameters can be used to generate arbitrarily large disordered configurations, which is absolutely essential for the convergence of the transport calculations \cite{Pro2}. Let us mention that the analysis developed in this study is not specific in any respect to the silicon crystal and, in principle, any other crystal can be analyzed in a similar fashion. We have strong reasons to believe that, apart from quantitative differences, the conclusions of this analysis on other crystals will be very similar. The implications of all these is the possibility to catalog the thermally-disordered configurations of the atoms that make up crystals, very much like we catalog their symmetries or the electron/phonon bands. 

\vspace{0.2cm}

The remaining of the paper is organized as follows. Section~\ref{Sec-CPMD} gives an overview of the {\it ab-initio} molecular dynamics (AIMD) methods and describes in detail the second generation CPMD method used in our study. Section~\ref{Sec-ErgodicSys} introduces the ergodic dynamical system $(\Omega,\GM,{\rm d}\PM)$ that defines the crystalline phase of silicon. Here, $\Omega$ is the the configuration space on which the space group $\GM$ acts by homeomorphisms. ${\rm d}\PM$ is the Gibbs measure over the configuration space, which must be ergodic w.r.t. the $\GM$-action \cite{Ruelle1969,Dobrushin1989}. The goal of section~\ref{Sec-ErgodicSys} is to define $(\Omega,\GM,{\rm d}\PM)$ theoretically, using the orbits provided by our CPMD simulations. Sections~\ref{Sec-ConfigSpace} and \ref{Sec-HGibbsMeasure} contain the concrete and detailed analyses of $\Omega$ and ${\rm d}\PM$ for the silicon crystal, respectively, while section~\ref{Sec-QGibbsMeasure} generates the full representation of the Gibbs measure. The last section concludes with advices on how to use the data generated by our study and how to generate similar data for more complex materials.

\section{Ab-Initio Molecular Dynamics}
\label{Sec-CPMD}

All of the nuclear trajectories were computed using the semi-classical AIMD, where the nuclei are treated classical while the interactions between them were determined ``on-the-fly'' by means of quantum mechanical density functional theory (DFT) calculations \cite{TDK2014,Parrinello1997}. 
The underlying basis of AIMD is the so-called Born-Oppenheimer (BO) approximation that, as we will show in the following, corresponds to the exact factorization of the electron-nuclear wave function, where the electron-nuclear coupling operator, which account for the electron-nuclear correlation, is neglected. In this section we briefly review the principles of these methods and present in more details the more recent, but already established \cite{CPMD2,TDK2014}, second generation AIMDs, which are employed in the present work.

\subsection{Exact Factorization of the Electron-Nuclear Wave Function}

Let us begin by considering the non-relativistic Hamiltonian describing the system of interacting electrons and nuclei, which, in the absence of an external electric field, reads as $\hat{H} = \hat{T}_N(\bm{R}) +  \hat{{H}}_{BO}(\bm{r}; \bm{R})$. Here, $\bm{r}=\{ \bm{r}_1, ..., \bm{r}_{N_{e}} \}$ and $\bm{R}=\{ \bm{R}_1, ..., \bm{R}_{N_{N}} \}$ are the coordinates of all $N_{e}$ electrons and $N_{N}$ nuclei, while $\hat{T}_N(\bm{R})$ corresponds to the nuclear kinetic energy operator and 
\begin{equation}
  \hat{{H}}_{BO}(\bm{r}; \bm{R}) = \hat{T}_e(\bm{r}) + \hat{V}_{ee}(\bm{r}) + \hat{V}_{eN}(\bm{r}, \bm{R}) + \hat{V}_{NN}(\bm{R})
\end{equation}
to the BO Hamiltonian, respectively. The latter consists not only of the electronic kinetic energy of operator $\hat{T}_e(\bm{r})$, but also of the various Coulomb operators describing the interactions among the electrons, between the electrons and nuclei, as well as among the nuclei. The nuclear Coulomb operator $\hat{V}_{NN}(\bm{R})$ is independent of $\bm{r}$ and represents a constant shift to the electronic energy. 

\vspace{0.2cm}

As proven by R. Hunter \cite{Hunter1975}, the full electron-nuclear wave function $\Psi(\bm{r}, \bm{R})$ can be exactly factorized as a product of a conditional probability distribution amplitude $\Phi(\bm{r}; \bm{R})$ for a given fixed nuclear configuration times a marginal probability amplitude $\chi(\bm{R})$, i.e. 
\begin{equation}
  \Psi(\bm{r}, \bm{R}) = \Phi(\bm{r}; \bm{R}) \chi(\bm{R}), 
\end{equation}
where $\Phi(\bm{r}; \bm{R})$ is the electronic and $\chi(\bm{R})$ the nuclear wave function. 


\vspace{0.2cm}

The non-adiabatic equations determining $\Phi(\bm{r}; \bm{R})$ and $\chi(\bm{R})$ are obtained by variationally optimizing the quantum expectation value of $\hat{H}$ w.r.t. the two wave functions, where the partial normalization constraints 
\begin{subequations}
\begin{eqnarray}
  \int d\bm{r} \, \left| \Phi(\bm{r}; \bm{R}) \right|^2 &=& 1 \\
    \int d\bm{R} \, \left| \chi(\bm{R}) \right|^2 &=& 1
\end{eqnarray}
\end{subequations}  
are imposed by means of Lagrangian multipliers \cite{Gidopoulos2014}. After some straightforward algebra, the eventual equations can be written as 
\begin{subequations}
\begin{eqnarray}
  \left[ \hat{{H}}_{BO}(\bm{r}; \bm{R}) + \hat{U}_{eN}(\bm{r}, \bm{R}) \right] \Phi(\bm{r}; \bm{R}) &=& \varepsilon(\bm{R})  \Phi(\bm{r}; \bm{R}) \label{ElecSE}\\
  \left[ \sum_{I=1}^{N_{N}} \frac{\left( -i \hbar \nabla_I + \bm{A}_{I}(\bm{R}) \right)^2}{2 M_I} + \varepsilon(\bm{R}) \right] \chi(\bm{R}) &=& E \chi(\bm{R}) \label{NuclSE}, 
\end{eqnarray}
\end{subequations}
where 
\begin{eqnarray}
  \hat{U}_{eN}(\bm{r}, \bm{R}) &=& \sum_{I=1}^{N_{N}} \frac{1}{M_I} \Biggl[ \frac{ \left( -i \hbar \nabla_I - \bm{A}_{I}(\bm{R}) \right)^2}{2} +  \left( \frac{\chi^*(\bm{R}) \left( -i \nabla_I \chi(\bm{R}) \right)}{\left| \chi(\bm{R}) \right|^2} + \bm{A}_{I}(\bm{R}) \right) \nonumber \\ 
  &\times& \left(-i \hbar \nabla_I - \bm{A}_{I}(\bm{R}) \right) \Biggr] \label {ENC}
\end{eqnarray}
is the electron-nuclear coupling operator \cite{Gross2014}. The scalar and vector potential terms within \eqref{ElecSE} and \eqref{NuclSE} are 
\begin{eqnarray}
  \varepsilon(\bm{R}) &=& \Biggl\langle \Phi(\bm{r}; \bm{R}) \Bigg| \hat{{H}}_{BO}(\bm{r}; \bm{R}) + \sum_{I=1}^{N_{N}} \frac{\left[ -i \hbar \nabla_{I} - \bm{A}_{I}(\bm{R}) \right]^2}{2M_I} \Bigg| \Phi(\bm{r}; \bm{R}) \Biggr\rangle_{\bm{r}} \\
    &=& \langle \Phi(\bm{r}; \bm{R}) | \hat{{H}}_{BO}(\bm{r}; \bm{R}) | \Phi(\bm{r}; \bm{R}) \rangle_{\bm{r}} + \sum_{I=1}^{N_{N}} \frac{1}{2M_I} \left[ \hbar^2 \langle \nabla_I \Phi(\bm{r}; \bm{R}) | \nabla_I \Phi(\bm{r}; \bm{R}) \rangle_{\bm{r}} - \bm{A}_{I}(\bm{R})^2 \right] \nonumber
\end{eqnarray}
and
\begin{equation}
  \bm{A}_{I} = \langle \Phi(\bm{r}; \bm{R}) | -i \nabla_I \Phi(\bm{r}; \bm{R}) \rangle_{\bm{r}}, 
\end{equation}
respectively, where $\langle \cdots \rangle_{\bm{r}}$ indicates integration over $\bm{r}$-space only. While the electronic energy $\varepsilon(\bm{R})$ is equivalent to the exact potential energy surface, the Berry-like vector potential $\bm{A}_{I}$ is also known as Berry connection \cite{Gross2010}. As already alluded to above, neglecting the electron-nuclear coupling operator $\hat{U}_{eN}(\bm{r}, \bm{R})$ corresponds to the well-known BO approximation \cite{BO1927}. 

\subsection{Second generation Car-Parrinello Molecular Dynamics}

Until recently, AIMD has mostly relied on two general methods: The original CPMD and the direct BOMD approach, each with its advantages and shortcomings. In BOMD, the total energy of a system, as determined by an arbitrary electronic structure method, is fully minimized in each MD time step, which renders this scheme computationally very demanding \cite{BOMD}. By contrast, the original CPMD technique obviates the rather time-consuming iterative energy minimization by considering the electronic degrees of freedom as classical time-dependent fields with a fictitious mass $\mu$ that are propagated by the use of Newtonian dynamics \cite{CP1985}. In order to keep the electronic and nuclear subsystems adiabatically separated, which causes the electrons to follow the nuclei very close to their instantaneous electronic ground state, $\mu$ has to be chosen sufficiently small \cite{GSB1991}. However, in CPMD the maximum permissible integration time step scales like $\sim \sqrt{\mu}$ and therefore has to be significantly smaller than that of BOMD, hence limiting the attainable simulation timescales \cite{BS1998}. 

\vspace{0.2cm}

The so-called second generation CPMD method combines the best of both approaches by retaining the large integration time steps of BOMD, while, at the same time, preserving the efficiency of CPMD \cite{CPMD2}. In this Car-Parrinello-like approach to BOMD, the original fictitious Newtonian dynamics of CPMD for the electronic degrees of freedom is substituted by an improved coupled electron-ion dynamics that keeps the electrons very close to the instantaneous ground-state without the necessity of an additional fictitious mass parameter. The superior efficiency of this method, which originates from the fact that only one preconditioned gradient computation is necessary per AIMD step, varies between one to two orders of magnitude, depending on the particular system \cite{TDK2014}. 

\vspace{0.2cm}

Within mean-field electronic structure methods, such as Hartree-Fock and Kohn-Sham DFT, the total energy $E \left[ \{ \psi_i \}; \bm{R} \right]$ is either a functional of the electronic wave function that is described by a set of occupied molecular orbitals (MOs) $\ket{\psi_i}$ or, equivalently, of the corresponding one-particle density operator $\rho = \sum_i {\ket{\psi_i} \bra{\psi_i}}$. The improved coupled electron-ion dynamics of second generation CPMD obeys the following equations of motion for the nuclear and electronic degrees of freedom: 
\begin{subequations}
\begin{eqnarray}
M_{I} \ddot{\bm{R}}_I &=& -\nabla_{{\bm{R}}_{I}} \left[ \left. \min_{ \{ \psi_i \}} \, E \left[ \{ \psi_i \}; \bm{R}_{I} \right] \right|_{\{ \braket{\psi_i}{\psi_j} = \delta_{ij} \}} \right] \\
&=& - \frac{\partial E}{\partial \bm{R}_{I}} + \sum_{i,j} \Lambda_{ij} \frac{\partial}{\partial \bm{R}_{I}} \braket{\psi_i}{\psi_j} \nonumber \\
&-& 2 \sum_{i} \frac{\partial \bra{\psi_i}}{\partial \bm{R}_I} \left[ \frac{\partial E \left[ \{ \psi_i \}; \bm{R}_{I} \right]}{\partial \bra{\psi_i}} - \sum_j \Lambda_{ij} \ket{\psi_j} \right] \nonumber \label{NuclEOM} \\
\frac{d^2}{d \tau^2} \ket{\psi_i (\bm{r}, \tau)} &=& - \frac{\partial E \left[ \{ \psi_i \}; \bm{R}_{I} \right]} {\partial \bra{\psi_i (\bm{r}, \tau)}} - \gamma \omega \frac{d}{d\tau} \ket{\psi_i (\bm{r}, \tau)} + \sum_j \Lambda_{ij} \ket{\psi_j(\bm{r}, \tau)}. \label{ElecEOM}
\end{eqnarray}
\end{subequations}
The former is the conventional nuclear equation of motion of BOMD consisting of Hellmann-Feynman, Pulay and non-self-consistent force contributions \cite{Hellmann1937,Feynman1939,Pulay1969,BendtZunger1983}, whereas the latter constitutes an universal oscillator equation 
as obtained by a nondimensionalization. The first term on the right-hand side in \eqref{ElecEOM} can be sought of as an ``electronic force'' to propagate $\ket{\psi_i}$ in dimensionless time $\tau$. The second term is an additional damping term to relax more quickly to the instantaneous electronic ground-state, where $\gamma$ is an appropriate samping coefficient and $\omega$ the undamped angular frequency of the universal oscillator. The final term derives from the constraint to fulfill the holonomic orthonormality condition $\braket{\psi_i}{\psi_j} = \delta_{ij}$, by using the Hermitian Lagrangian multiplier matrix $\bm{\Lambda}$. As can be seen in \eqref{ElecEOM}, not even a single diagonalization step, but just one ``electronic force'' calculation is required. In other words, the second generation CPMD method not only entirely bypasses the necessity of a self-consistent field cycle, but also the alternative iterative wave function optimization. 

\vspace{0.2cm}

However, contrary to the evolution of the nuclei, for the short-term integration of the electronic degrees of freedom accuracy is crucial, which is why a highly accurate yet efficient propagation scheme is essential. As a consequence, the evolution of the MOs is conducted by extending the always-stable predictor-corrector integrator of Kolafa to the electronic structure problem \cite{KolafaASPC}. But, since this scheme was originally devised to deal with classical polarization, special attention must be paid to the fact that the holonomic orthonormality constraint, which is due to the fermionic nature of electrons that forces the wave function to be antisymmetric in order to comply with the Pauli exclusion principle, is always satisfied during the dynamics. 
For that purpose, first the predicted MOs at time $t_n$ are computed based on the electronic degrees of freedom from the K previous AIMD steps: 
\begin{equation}
\ket{\psi_i^p(t_n)} = \sum_{m}^{K} (-1)^{m+1} m \frac{\binom{2K}{K-m}}{\binom{2K-2}{K-1}} \rho(t_{n-m}) \ket{\psi_i(t_{n-1})}.
\end{equation}
This is to say that the predicted one-electron density operator $\rho^p(t_{n})$ is used as a projector onto to occupied subspace $\ket{\psi_i(t_{n-1})}$ of the previous AIMD step. In this way, we take advantage of the fact that $\rho^p(t_n)$ evolves much more smoothly than $\ket{\psi_i^p(t_{n})}$ and is therefore easier to predict. This is particularly true for metallic systems, where many states crowd the Fermi level. Yet, to minimize the error and to further reduce the deviation from the instantaneous electronic ground-state, $\ket{\psi_i^p(t_n)}$ is corrected by performing a single minimization step $\ket{\delta \psi_i^p(t_n)}$ along the preconditioned electronic gradient direction, as computed by the orthonormality conserving orbital transformation method \cite{OT}. Therefore, the modified corrector can be written as 
\begin{eqnarray}
\ket{\psi_i(t_n)} &=& \ket{\psi_i^p(t_n)} + \omega \left( \ket{\delta \psi_i^p(t_n)} - \ket{\psi_i^p(t_n)} \right), \nonumber \\
\text{with}~\omega &=& \frac{K}{2K-1}~\text{for}~K \ge 2.
\end{eqnarray}
The eventual predictor-corrector scheme leads to an electron dynamics that is rather accurate and time-reversible up to $\mathcal{O}(\Delta t^{2K-2})$, where $\Delta t$ is the discretized integration time step, while $\omega$ is chosen so as to guarantee a stable relaxation towards the minimum. In other words, the efficiency of the present second generation CPMD method stems from the fact that the instantaneous electronic ground state is very closely approached within just one electronic gradient calculation. We thus totally avoid the self-consistent field cycle and any expensive diagonalization steps, while remaining very close to the BO surface and, at the same time, $\Delta t$ can be chosen to be as large as in BOMD. 

\vspace{0.2cm}

However, in spite of the close proximity of the propagated MOs to the instantaneous electronic ground state, the nuclear dynamics is slightly dissipative, most likely because the employed predictor-corrector scheme is not symplectic. Nevertheless, presuming that the energy is exponentially decaying, which had been shown to be an excellent assumption \cite{TDK2014,TDK2009}, it is possible to rigorously correct for the dissipation by modeling the nuclear forces of second generation CPMD $\bm{F}^{CP}_I = - \nabla_{\bm{R}_I} E \left[ \{ \psi_i \}; \bm{R}_{I} \right]$ by 
\begin{equation}
\bm{F}_I^{CP} = \bm{F}_I^{BO}-\gamma_D M_I \dot{\bm{R}}_I, 
\end{equation}
where $\bm{F}_I^{BO}$ are the exact, but inherently unknown BO forces and $\gamma_D$ is an intrinsic, yet to be determined friction coefficient to mimic the dissipation. The presence of damping immediately suggests a canonical sampling of the Boltzmann distribution by means of the following modified Langevin-type equation: 
\begin{subequations}\label{Eq-Langevin}
\begin{eqnarray}
M_I \ddot{\bm{R}}_I &=& \underbrace{\bm{F}_I^{BO} - \gamma_D M_I \dot{\bm{R}}_I} + \bm{\Xi}_I^{D} \\
&=& \qquad \; \; \bm{F}_I^{CP}+ \bm{\Xi}_I^{D}, 
\end{eqnarray}
\end{subequations}
where $\bm{\Xi}_I^{D}$ is an additive white noise term, which must obey the fluctuation-dissipation theorem $\langle \bm{\Xi}_I^D(0) \bm{\Xi}_I^D(t) \rangle = 2 \gamma_D M_I k_B T \delta(t)$ in order to guarantee an accurate sampling of the Boltzmann distribution \cite{Kubo1966}. 

\vspace{0.2cm}

This is to say that if one knew the unknown value of $\gamma_D$ it would nevertheless be possible to ensure an exact canonical sampling of the Boltzmann distribution in spite of the dissipation. Fortunately, the inherent value of $\gamma_D$ does not need to be known \textit{a priori}, but can be bootstrapped so as to generate the correct average temperature \cite{CPMD2}, as measured by the equipartition theorem 
\begin{equation} \label{EPT}
  \left< \frac{1}{2} M_I \dot{\bm{R}}_I^2 \right> = \frac{3}{2} k_B T. 
\end{equation}
More precisely, in order to determine the hitherto unknown value of $\gamma_D$, we perform a preliminary simulation in which we vary $\gamma_D$ on-the-fly using a Berendsen-like algorithm until \eqref{EPT} is eventually satisfied \cite{TDK2009}. 

\subsection{Computational Details}

Our models of crystalline and liquid silicon consisted of 216, 512 and 1000 Si atoms in a cubic simulation box with periodic boundary conditions. Specifically, for each system size considered, 10 simulations with different temperatures between 300~K and 3000~K, which were equally separated by 300~K each, have been conducted. As already mentioned, all of our calculations were performed in the canonical NVT ensemble using the second generation CPMD method \cite{CPMD2,TDK2014}. Throughout, the experimental density of crystalline silicon was assumed, which, at ambient conditions, is semiconducting and fourfold coordinated. However, upon melting, which happens at $\sim$1685~K, the density of silicon increases by 10\% and its structure transitions into a unusually low coordinated metallic liquid \cite{Parrinello1997}. 
The modified Langevin equation was integrated based on the algorithm of Ricci and Ciccotti \cite{RicciCiccotti}, with a discretized time step of 1.0~fs. The friction coefficient $\gamma_D$, needed to satisfy the equipartition theorem, turned out to be in the range of 10$^{-4}$fs$^{-1}$. 
In each run, we carefully equilibrated the system for 250~ps before accumulating statistics during additional 1.25~ns, resulting in a total AIMD simulation time of 45~ns. 

\vspace{0.2cm}

All simulations were performed at the DFT level using the mixed Gaussian and plane wave code CP2K/\textit{Quickstep} \cite{Quickstep}. In this approach, the Kohn-Sham orbitals are expanded in contracted Gaussians functions, while the electronic charge density is represented by plane waves \cite{GPW}. For the former, a dimer-optimized basis set, each with two s- and p-type Gaussian exponents \cite{Zijlstra}, is employed, while for the latter a density cutoff of 100~Ry is used. The unknown exact exchange and correlation potential is substituted by the local-density approximation \cite{GTH}, whereas the interactions between the valence electrons and the ionic cores are described by separable norm-conserving Goedecker-Teter-Hutter pseudopotentials \cite{GTH,PP}.  For the sake of simplicity, the first Brillouin zone of the supercell is sampled at the $\Gamma$-point only. 


\section{The Ergodic Dynamical System Characterizing the Atomic Configurations}
\label{Sec-ErgodicSys}

\begin{figure}
\center
  \includegraphics[width=0.7\textwidth]{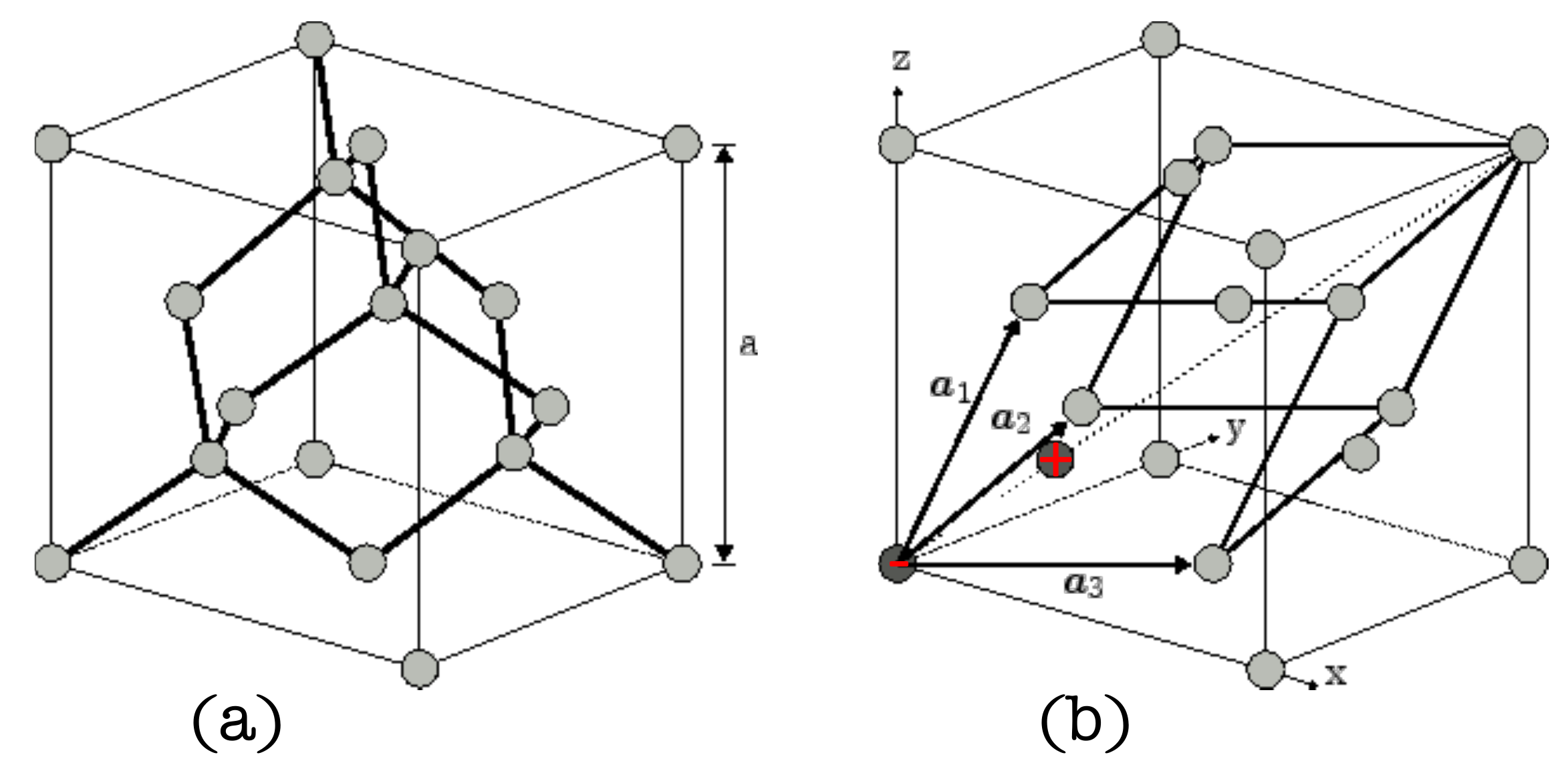}\\
  \caption{\small (Adopted from http://www.iue.tuwien.ac.at/phd/ungersboeck/node27.html) Silicon crystalizes in a diamond cubic lattice (Fd-3m), whose conventional unit cell is shown in panel (a). This cubic unit cell is symmetric to the full space group and contains eight Si atoms. The diamond cubic lattice results from the inter-penetration of two face-centered cubic (fcc) lattices. The fcc lattice can be generated by translating a primitive cell that contains just one atom. Hence, silicon's diamond cubic lattice can be generated by translating the same primitive cell, but with one additional Si atom inside it. This primitive cell is shown in panel (b), together with its two atoms  ($\pm$) and the generating primitive vectors $\vec a_1=\tfrac{a}{2}(\hat z + \hat y)$, $\vec a_2=\tfrac{a}{2}(\hat x + \hat z)$ and $\vec a_3=\tfrac{a}{2}(\hat y + \hat x)$, with $a=5.431$~\AA. It is more convenient to shift the primitive cell such that its center coincides with the midpoint between the $\pm 1$ atoms.}
 \label{SiPrimitiveCell}
\end{figure}

Silicon crystallizes in the diamond cubic crystal structure, which can be visualized as two interpenetrating face-centered cubic lattices. Its space group is $\GM=Fd\bar 3$m \cite{Singh93,IntTables}. The conventional unit cell is shown in panel (a) of Fig.~\ref{SiPrimitiveCell}. It is of cubic shape with a lattice constant at zero temperature of $a=5.431 \AA$ and contains a total of eight Si atoms. Throughout, all distances will be reported in Angstroms. The primitive cell, which is the smallest domain that generates the entire lattice when translated by integer multiples of the primitive vectors $\vec a_i$, $i=1,2,3$, is shown in Fig.~\ref{SiPrimitiveCell}(b). There are two atoms per primitive cell and these two atoms will be labeled by the index $\alpha=\pm 1$. The zero-temperature equilibrium positions of the atoms located in the translated primitive cell labeled by $\bm n \in \ZM^3$ are:
\begin{equation}\label{Eq-LatticePoints}
{\bm x}_{\bm n}^\alpha = n_1 \vec a_1 + n_2 \vec a_2 + n_3 \vec a_3 + \tfrac{\alpha}{8}(\vec a_1 + \vec a_2 + \vec a_3), \quad \alpha = \pm 1.
\end{equation} 
These points form the ideal silicon lattice:
\begin{equation}
\Ll = \Big \{ \bm x_{\bm n}^\alpha, \ \bm n \in \ZM^3, \ \alpha=\pm 1 \Big \} \subset \RM^3,
\end{equation}
which is left invariant by the space group $\GM$. The action of the latter on $\bm x \in \Ll$ will be denoted by $\bm g \cdot \bm x$. Let us recall that the asymmetric unit cell of the diamond cubic structure contains a single atom, which means that, for any fixed $x\in \Ll$,  $\Ll=\big \{ \bm g \cdot \bm x, \ \bm g \in \GM\big \}$. In other words, the entire lattice can be reconstructed from just one atom by applying all space group transformations \cite{Singh93,IntTables}.

\vspace{0.2cm}

At finite temperatures, the atoms are in a constant thermal motion and, in an instantaneous snapshot, they will appear in a disordered configuration. In this work, the statistical mechanics associated to this dynamics is explored using the second generation CPMD algorithm described in section~\ref{Sec-CPMD}. To avoid constant comparison between the physical reality and the output of our CPMD simulations, which is not the purpose of our work here, we will take the view that CPMD provides the ``physical reality''. Henceforth, the goal of this section is to define the space $\Omega$ of the atomic configurations and the action $\tau$ of the space group on $\Omega$, using only the output of the present CPMD simulations, namely, the temporal orbits of the atoms at various temperatures, which are inherently finite. Together with the probability measure ${\rm d} \mathbb P$, provided by the Gibbs measure for the atomic degrees of freedom, $(\Omega,\GM,\tau,{\rm d} \mathbb P)$ define an ergodic dynamical system that characterizes entirely the crystalline phase of silicon. Let us be clear that the goal of this section is to give a formal definition of this dynamical system, leaving the quantitative analysis for the subsequent sections.

\subsection{The Configuration Space Defined}

The following facts regarding the CPMD simulations of crystalline phases will play an essential role in the analysis to follow:
\begin{enumerate}[\rm (f1)]

\item The atoms are initially placed at the zero-temperature equilibrium positions.

\item The atoms are given initial velocities drawn from the Maxwell distribution corresponding to a particular temperature.

\item The macroscopic properties of the crystal, as computed from the present CPMD simulations, do not fluctuate from one set of initial conditions to another, provided that 1) the crystal size is sufficiently large, 2) the initial velocities sample well the Maxwell distribution, and 3) the temporal orbits are long enough.

\end{enumerate}
The observation stated in (f3) automatically implies that each (infinite) temporal orbit densely fills the configuration space of the silicon crystal. In other words, the temporal dynamics is ergodic.

\vspace{0.2cm}

In the CPMD simulations, one can follow the motion of each atom. As such, the atoms can be labeled by their initial positions, which, according to (f1), are fully specified by the primitive cell index $\bm n \in \ZM^3$ and by the atomic index $\alpha = \pm 1$, or simply by $\bm x \in \Ll$. In the thermodynamic limit, the dynamics of the atoms can be represented as the motion of a point $\omega$ in the topological space $\prod_{\bm x \in \Ll} \RM^{3}$. Indeed, if $\vec R_{\bm x}(t)$ represents the instantaneous coordinates of the atom located initially at $\bm x \in \Ll$, then the dynamics of the crystal can be represented as:
\begin{equation}\label{Eq-Orbit}
\RM_+ \ni t \mapsto \omega(t) \in \prod_{\bm x \in \Ll} \RM^{3}, \quad \omega_{\bm x}(t) = \vec R_{\bm x}(t)-\bm x \in \RM^3.
\end{equation}
We note that the subtraction of the equilibrium positions will be essential for making the configuration space into a compact set. The points of  $\prod_{\bm x \in \Ll} \RM^{3}$ will be represented as infinite sequences: 
\begin{equation}
\omega = \{\omega_{\bm x}\}_{\bm x \in \Ll} \in \, \prod_{\bm x \in \Ll} \RM^{3}.
\end{equation} 
When we need to specify the point $\bm x$ precisely, we proceed as in \eqref{Eq-LatticePoints} and use its corresponding primitive cell index $\bm n \in \ZM$ and atomic index $\alpha = \pm 1$. There is a standard action of $\GM$ on this space:
\begin{equation}
\tau_{\bm g} \omega = \tau_{\bm g}\{\omega_{\bm x}\}_{\bm x \in \Ll} = \{\omega_{\bm g \cdot \bm x}\}_{\bm x \in \Ll}, \quad {\bm g}\in \GM.
\end{equation}

\vspace{0.2cm}

Regarding the topology of $\prod_{\bm x \in \Ll} \RM^{3}$, we recall that, if: 
\begin{equation}
{\rm P}_{\bm x}: \prod_{\bm x \in \Ll} \RM^{3} \rightarrow \RM^{3}, \quad {\rm P}_{\bm x} \omega = \omega_{\bm x}, \quad \bm x \in \Ll,
\end{equation}
represent the projections onto the $\bm x$-components, then the topology of the product space $\prod_{\bm x \in \Ll} \RM^{3}$ is given by the coarsest topology that makes all ${\rm P}_{\bm x}$ into continuous maps. In other words, a map $f$ with values in this space is continuous if and only if ${\rm P}_{\bm x} \circ f$ are all continuous. This characterization will be used often in our analysis. We mention the useful identity: \begin{equation}
P_{\bm g \cdot \bm x} = P_{\bm x} \circ \tau_g, \quad {\bm x} \in \Ll, \quad {\bm g} \in \GM.
\end{equation}
 
 \vspace{0.2cm}
 
The space we just described above is too big and, besides the convenient representation of the atomic configurations, it does not bring anything else useful. The configuration space $\Omega$ we are looking for is a subset of it, which needs to be compact and invariant to the full space group. Formally, we can invoke the ergodicity of the time evolution and define the configuration space as the closure of one temporal orbit, in the product topology of $\prod_{\bm x \in \Ll} \RM^{3}$ described above. More precisely:
\begin{equation}\label{Eq-ConfigSpace}
\Omega = \overline{ \big \{\omega(t), \ t \in \RM_+ \big \} } \subset \prod_{\bm x \in \Ll} \RM^{3}.
\end{equation}
Although formal, this definition is particularly useful in the present context because our CPMD simulations provide relatively long orbits, which can then be used, constructively, to explore the configuration space of the crystal and derive approximate, yet quantitative representations of $\Omega$.  

\vspace{0.2cm}

We need to alert the reader that definition \eqref{Eq-ConfigSpace} is correct only for our CPMD crystal. For a real physical crystal, at any finite temperature, there is a small but nevertheless non-zero probability for two atoms to exchange positions. Since we labeled the atoms, such process leads in \eqref{Eq-ConfigSpace} to two distinct configurations, which is clearly a spurious effect since the atoms are indistinguishable (a well known consequence of such over-counting is the Gibbs paradox). To avoid such difficulties, one should consider the time evolution of the un-labeled point-pattern defined by the atomic positions, which live in the space of point-patterns, as explained in \cite{Bel2015}. For the CPMD orbits considered in our study, the atoms never exchange positions, hence both approaches lead to the same configuration space. 

\subsection{The Topological Dynamical System}

Besides the dynamical system associated with the temporal evolution, there is a discrete dynamical system associated with the space group of the crystal, which plays an essential role in the theory of homogeneous phases \cite{Pro2}. Our goal here is to define this dynamical system based entirely on the temporal orbits. For this, we consider the action: 
\begin{equation}
\vec R_{\bm x}(t) \mapsto \tau'_{\bm g}\vec R_{\bm x}(t), \quad \bm g \in \GM, \quad \bm x \in \Ll, \quad t \in \RM_+,
\end{equation} 
of the space group $\GM$ on the individual temporal orbits of the atoms, where the new orbits $\tau'_{\bm g}\vec R_{\bm x}$ are defined by integrating the modified Langevin equation \eqref{Eq-Langevin} with the initial conditions:
\begin{equation}
\tau'_{\bm g}\vec R_{\bm x}(0) = \vec R_{\bm x}(0)=\bm x, \quad \left . \frac{{\rm d} \tau'_{\bm g}\vec R_{\bm x}}{{\rm d}t} \right |_{t=0} = \left .\frac{{\rm d} \vec R_{\bm g \cdot \bm x}}{{\rm d}t} \right |_{t=0}.
\end{equation}
At its turn, this will induce an action: 
\begin{equation}
\omega(t) \rightarrow \tau_{\bm g} \omega(t), \quad \bm g \in \GM,
\end{equation} 
and, since we subtract the initial position in the definition \eqref{Eq-Orbit} of $\omega$, this action relates to the standard action of $\GM$ on the larger space $\prod_{\bm x \in \Ll} \RM^{3}$:
\begin{equation}
\tau_{\bm g} \omega(t) = \tau_{\bm g} \{ \omega_{\bm x}(t)\}_{\bm x \in \Ll} = \{ \omega_{\bm g \cdot \bm x}(t)\}_{\bm x \in \Ll}, \quad \bm g \in \GM.
\end{equation}
Now, note that the shifted initial velocities still obey the Maxwell distribution, and it is here where observation (f3) becomes essential. Indeed, (f3) can hold if and only if $\tau_{\bm g} \omega(t)$ densely fills the configuration space, or in mathematical terms:  
\begin{equation}
\overline{ \big \{\tau_{\bm g}\omega(t), \ t \in \RM_+ \big \} } = \overline{ \big \{\omega(t), \ t \in \RM_+ \big \} }.
\end{equation} 
The conclusion is that the configuration space $\Omega$, as formally defined in \eqref{Eq-ConfigSpace}, is indeed invariant relative to the standard action of $\GM$ on $\prod_{\bm x \in \Ll} \RM^{3}$, hence we can restrict this action on $\Omega$ itself and define:
\begin{equation}
(\GM, \Omega) \ni (\bm g, \omega) \rightarrow \tau_{\bm g} \, \omega = \tau_{\bm g}\{\omega_{\bm x} \}_{\bm x \in \Ll} = \{\omega_{\bm g \cdot \bm x} \}_{\bm x \in \Ll}.
\end{equation}
These are continuous invertible maps over $\Omega$, hence we can endow the configuration space with the structure of a topological dynamical system $(\Omega,\GM,\tau)$. 

\subsection{The Gibbs Measure}

The configuration space can be also endowed with a probability measure provided by the Gibbs measure:
\begin{equation}
{\rm d} \PM(\omega) = \lim_{N\rightarrow \infty} Z_N e^{-\beta V_N(\omega_N)}{\rm d}\omega_N,
\end{equation}
whenever this limit exists. Above, $N$ refers to a finite crystal containing $N^3$ primitive cells and $V_N$ is the inter-atomic potentials as derived from our CPMD simulations on the finite crystal. Now, upon heating, a crystal can undergo structural phase transitions though this is not the case for silicon, which remains in a single pure crystalline phase all the way to melting. A very general conjecture, whose precise statement can be found in \cite[Ch.~6]{Ruelle1969}, states that, for pure homogenous thermodynamic phases, the Gibbs measure must be invariant and ergodic w.r.t. the space translations. For a pure crystalline phase, these two properties must hold for the full space group. Let us state explicitly that invariance means:
\begin{equation}
\PM (\tau_{\bm g} \Delta) = \PM(\Delta), \quad \forall \ \bm g \in \GM, \quad \Delta \subseteq \Omega,
\end{equation}
or simply $\PM \circ \tau_{\bm g} = \PM$ for all $\bm g \in \GM$. The conclusion is that the statistical mechanics of the Si atoms in the crystalline phase is fully encoded in the measure-preserving ergodic dynamical system $(\Omega,\GM,\tau,{\rm d} \PM)$. This dynamical system is at the core of homogeneous crystal theory and is essential for the computation of the macroscopic properties of the realistic crystals. One of the main goals of our work is to quantify the Gibbs measure.

\section{Mapping the Configuration Space}
\label{Sec-ConfigSpace}

The goal of this section is to analyze and quantify the configuration space. Approximate representations of it are in principle computable from the atomic orbits, but its structure may be very complex. To comprehend this structure, we propose a hierarchy of characterizations that will enable us to generate a tower of increasingly accurate representations for $\Omega$. In essence, we are going to piece together $\Omega$ from projections on various coordinate planes.

\subsection{Temporal Atomic Orbits: Projecting on One Atom}
\label{Sec-TemOrb1}

\begin{figure}[H]
\center
  \includegraphics[width=\textwidth]{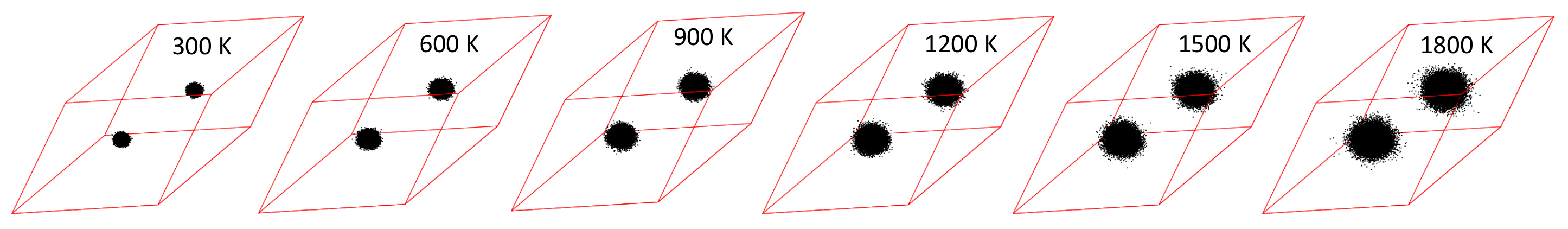}\\
  \caption{\small Three dimensional renderings of the orbits. They correspond to the two Si atoms that start their motion in the first primitive cell. The orbits have been mapped at different temperatures ranging from $T=300\ K$ to $1800\ K$, as indicated in each panel. The primitive cell is also shown in red.}
 \label{TempDepOrbits}
\end{figure}

 Typical temporal orbits of the two Si atoms in the first primitive cell, moving under the modified Langevin equation \eqref{Eq-Langevin}, are reported in Fig.~\ref{TempDepOrbits}. The orbits have been mapped for an array of temperatures, ranging from $300\ K$ to $1800\ K$. Note that the highest temperature is above the real melting temperature of crystalline silicon and, as we shall see shortly, not all orbits look as in Fig.~\ref{TempDepOrbits} at $1800\ K$. 
 
 The important feature revealed by this type of data is that, for temperatures up to $1500\ K$, the atoms remain confined in well defined neighborhoods of the equilibrium positions. These neighborhoods remain well separated from each other and from the facets of the primitive cell. Let us then introduce the closed topological spaces:
\begin{equation}
\Omega_{\bm x} = \overline{ \{ {\rm P}_{\bm x} \omega(t), \ t \in \RM_+\} } = \overline{ \{ \omega_{\bm x}(t), \ t \in \RM_+\} } \subset \RM^3, \quad \bm x \in \Ll,
\end{equation}
which have the approximate representations seen in Fig.~\ref{TempDepOrbits}. In particular, they are compact sets. Now, since ${\rm P}_{\bm x}$ are continuous maps:
\begin{equation}
{\rm P}_{\bm x} \Big (\, \overline{ \big \{\omega(t), \ t \in \RM_+ \big \} } \, \Big ) = \overline{\big  \{ {\rm P}_{\bm x} \omega(t), \ t \in \RM_+ \big \} },
\end{equation}
which gives us the first useful characterization of the configuration space:
\begin{equation}\label{Eq-Char1}
{\rm P}_{\bm x}(\Omega) = \Omega_{\bm x}, \quad \forall \ \bm x  \in \Ll.
\end{equation}
Furthermore, since the whole lattice $\Ll$ can be generated from any $\bm x \in \Ll$ by applying symmetries from the space group, the invariance of $\Omega$ w.r.t. the space group automatically implies that all $\Omega_{\bm x}$ are identical:
\begin{equation}
{\rm P}_{\bm g \cdot \bm x}(\Omega) = {\rm P}_{\bm x}(\bm g \cdot \Omega) = {\rm P}_{\bm x}(\Omega), \quad \forall \, \bm g \in \GM,
\end{equation} 
and will be denoted by the same symbol $\Omega_0$. One important conclusion from this and \eqref{Eq-Char1}, namely:
\begin{equation}
\Omega \subseteq \prod_{\bm x \in \Ll} \Omega_{\bm x} = \big ( \Omega_0 \big )^{\Ll}.
\end{equation}
The space on the right is an infinite product of compact spaces and Tychonoff's theorem assures us that the result is compact too. As a closed sub-set of a compact set, the configuration space $\Omega$, as defined in \eqref{Eq-ConfigSpace}, ought to be compact. This is one of the important qualitative properties of $\Omega$ we were looking for.

\begin{figure}[H]
\center
  \includegraphics[width=\textwidth]{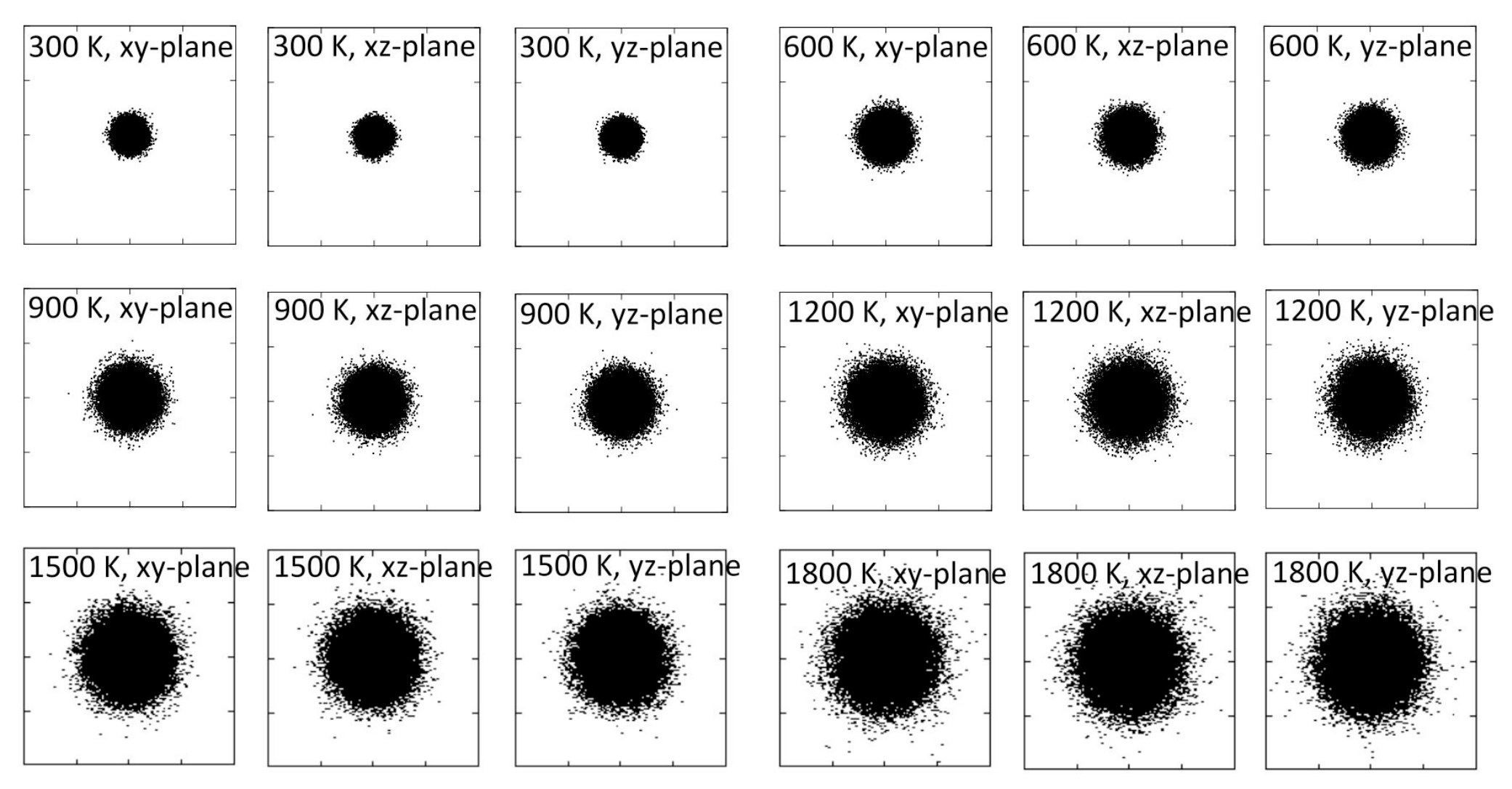}\\
  \caption{\small Projections of the orbit of one Si atom on different coordinate planes. For temperatures up to $1500\ K$, such projections are typical for all atoms included in the simulations. For $1800\ K$, there is a small fraction of atoms that leave the vicinity of their initial starting point. The boxes are all equal to $[-1,1]\times [-1,1]$ (recall that the unit for distance is the Angstrom).}
 \label{Fig-OneAtomOrbit}
\end{figure}

The set $\Omega_0$ is further examined in Fig.~\ref{Fig-OneAtomOrbit}, by projecting the orbits on various planes. This gives us a more detailed picture of this set, which reveals its somewhat surprising spherical symmetry. The set seems to be extremely well approximated by a solid ball. Quantitative assessments will be provided in the following sections. Let us also point out that $\Omega_0$ is centered at the origin, which indicates that the time average $\langle r_{\bm x}(t) \rangle \approx \bm x$. This, of course, can be verified quantitatively and we find the statement to be true with at least two significant digits of precision.

\subsection{Temporal Atomic Orbits: Projecting on Two Atoms}

To further investigate the shape of the configuration space, we consider projections on the coordinates of two atoms:
\begin{equation}
{\rm P}_{\bm x_1,\bm x_2} \omega = \big ( \omega_{\bm x_1},\omega_{\bm x_2} \big ) \in \RM^3 \times \RM^3, \quad \bm x_1 \neq \bm x_2,
\end{equation}
and investigate the sets ${\rm P}_{\bm x_1,\bm x_2}(\Omega)$. They live in a 6-dimensional space and, to visualize them, we will project the orbits on different planes, such as:
\begin{equation}
\Big ( \omega_{\bm x_1}(t)_{u_1}, \omega_{\bm x_2}(t)_{u_2} \Big ), \quad u_1,u_2=x,y,z, \quad t \in \RM_+.
\end{equation}

\vspace{0.2cm}

In the ideal silicon crystal, all pairs of atoms can be organized in classes $\Pp_n$, $n=1,2,\ldots$, of first, second, etc., near neighbors. For all pairs from the class $\Pp_n$ of $n^{\rm th}$-near neighbors, the atoms are separated by the same distance $d_n$ and $d_1<d_2<\ldots$. Let us recall that, for the ideal silicon lattice:
\begin{itemize}
\item $d_1=\frac{\sqrt{3}a}{4}$ and each Si atom has four $1^{\rm st}$-near neighboring atoms.
\item $d_2=\frac{\sqrt{2}a}{2}$ and each Si atom has twelve $2^{\rm nd}$-near neighboring atoms.
\item $d_3=\frac{\sqrt{11}a}{4}$ and each Si atom has twelve $3^{\rm rd}$-near neighboring atoms.
\item $d_4=a$ and each Si atom has six $4^{\rm th}$-near neighboring atoms.
\end{itemize}
Any two pairs of atoms from the same class $\Pp_n$, at least for $n=1,\ldots,4$, are connected by a space group transformation. In other words, the orbit $\{(\bm g \cdot \bm x_1,\bm g \cdot \bm x_2), \ \bm g \in \GM\}$ of any fixed pair from $\Pp_n$ generates the entire $\Pp_n$. This is important because, for a pair $(\bm x_1,\bm x_2) \in \Pp_n$, we have:
\begin{equation}\label{Eq-Omega1}
{\rm P}_{\bm g \cdot \bm x_1,\bm g \cdot \bm x_2}(\Omega) = {\rm P}_{\bm x_1,\bm x_2}(\bm g \cdot \Omega) = {\rm P}_{\bm x_1,\bm x_2}(\Omega), \quad \forall \, g \in \GM,
\end{equation}
hence all ${\rm P}_{\bm x_1,\bm x_2}(\Omega)$ are identical and will be denoted by $\Omega_n$ from now on. Obviously, $\Omega_n \subseteq \Omega_0 \times \Omega_0$.

\begin{figure}
\center
  \includegraphics[width=0.95\textwidth]{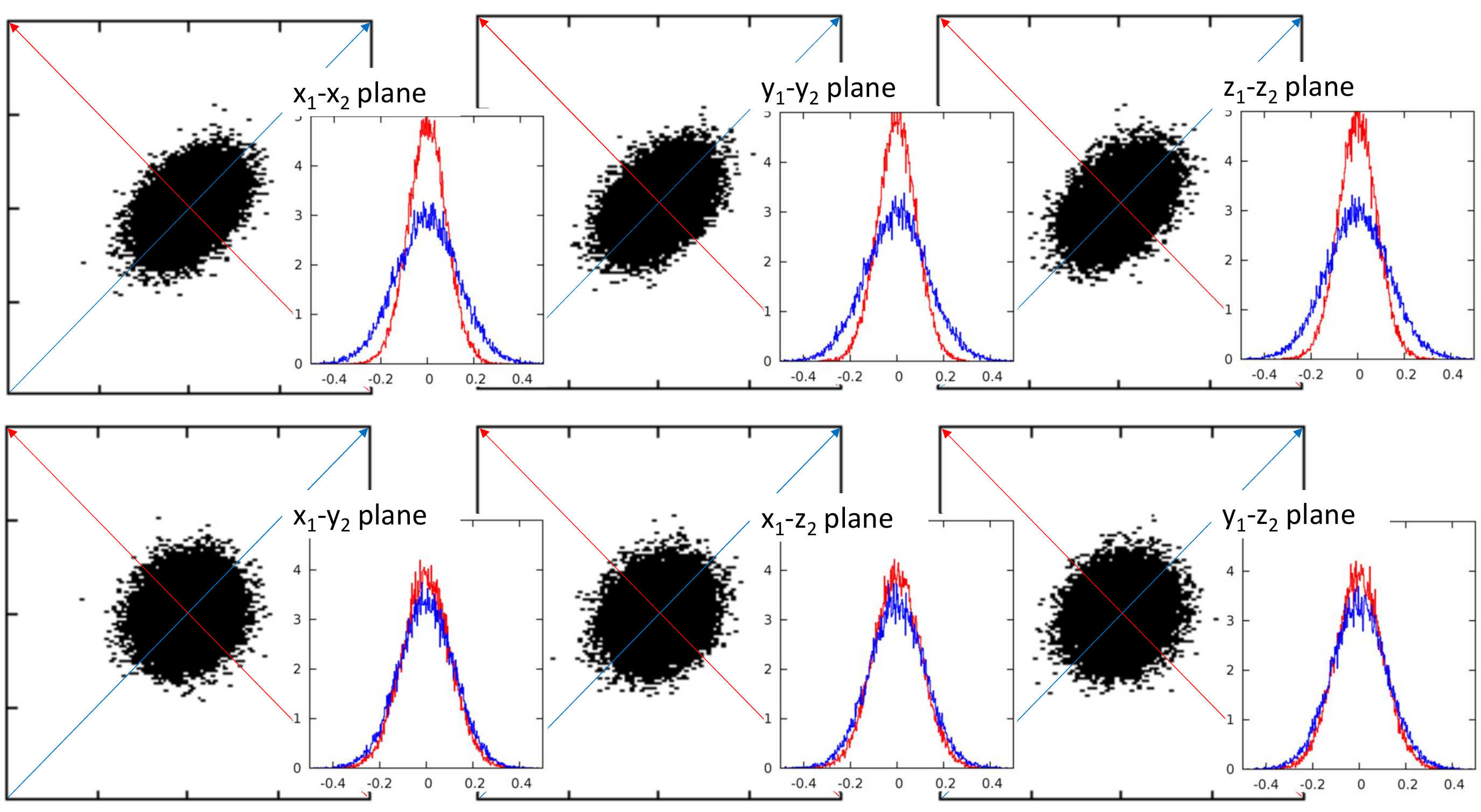}\\
  \caption{\small Projection of the orbit $ \omega(t)$ on coordinate planes determined by its $1^{\rm st}$-near neighboring atoms. More precisely, the panels show $\Big ( {\rm P}_{\bm x_1} \omega(t)_u,{\rm P}_{\bm x_2}\omega(t)_v \Big )$, where $\bm x_1=\bm x_{\bm 0}^{+1}$, $\bm x_2=\bm x_{\bm 0}^{-1}$ and $u=x_1,y_1,z_1$ and $v=x_2,y_2,z_2$, as specified on each panel. The size of the plotting boxes are all equal to $[-1,1] \times [-1,1]$ and the orbit is computed at $900\ K$. The insets show the density of the points when the data is projected on the first (blue) and second (red) diagonals of the coordinate systems in the main panels.}
 \label{Fig-900K1StPairProjOrbits}
\end{figure}

Any correlation between the motions of the two atoms will result in an anisotropy of the projected orbits. Such correlations are expected to be strong for $1^{\rm st}$-near neighboring atoms and to become weaker for further neighboring atoms. In fact, our analysis suggests that the correlations become negligible beyond the $4^{\rm th}$-near neighboring atoms. While we have analyzed all the temperatures appearing in the previous figures, the results and conclusions are very similar, hence we will only showcase the data at $900\ K$. 

\begin{figure}
\center
  \includegraphics[width=0.95\textwidth]{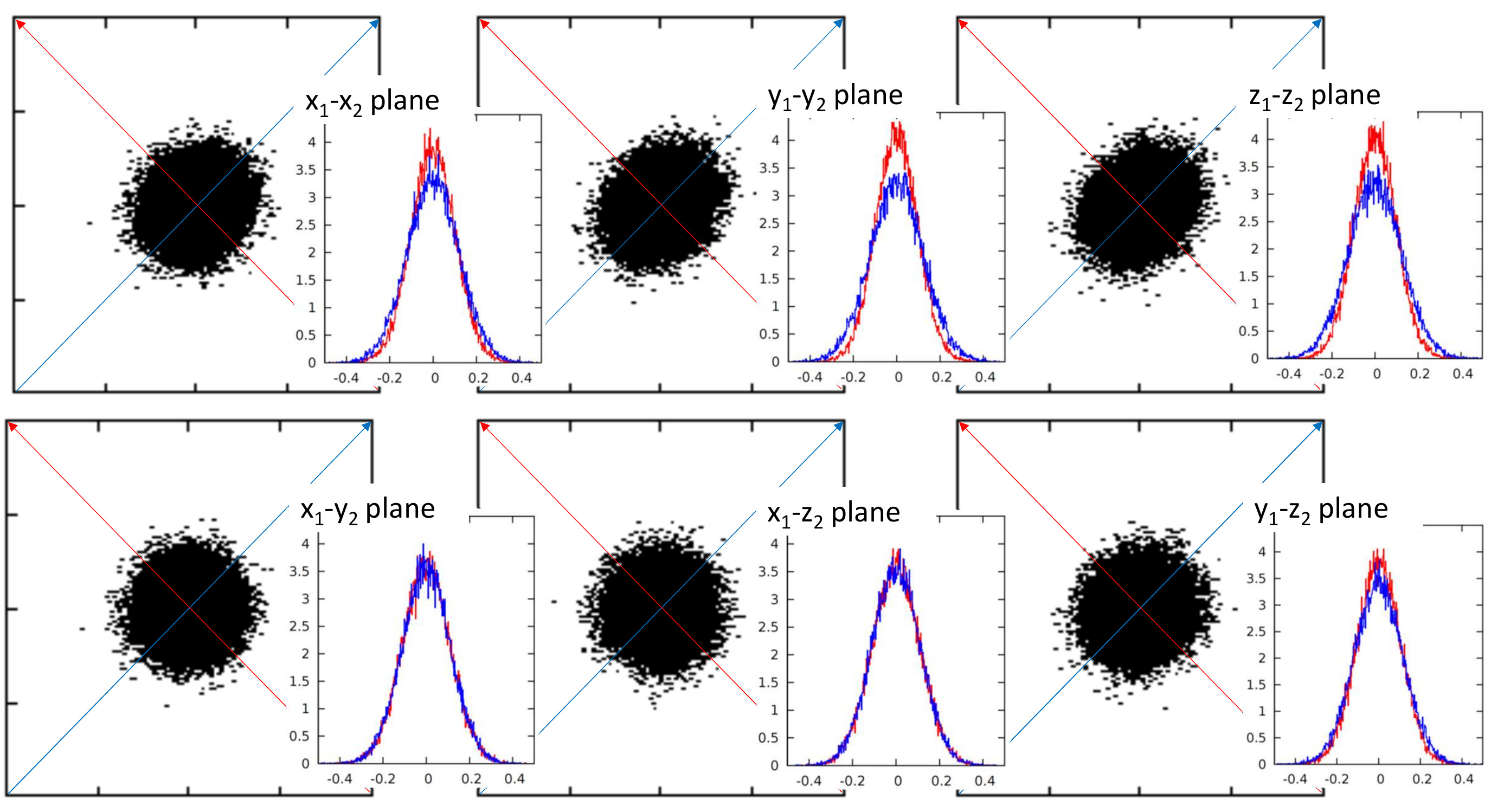}\\
  \caption{\small Same as Fig.~\ref{Fig-900K1StPairProjOrbits}, but for the pair of $2^{\rm nd}$-near neighbors $\bm x_1=\bm x_{(0,0,0)}^{-1}$ and $\bm x_2=\bm x_{(1,0,0)}^{-1}$.}
 \label{Fig-900K2NdPairProjOrbits}
\end{figure}

Fig.~\ref{Fig-900K1StPairProjOrbits} reports the results for one pair of $1^{\rm st}$-near neighboring atoms, whose projected orbit gives us a first representation of $\Omega_1$.  Examining the data, one can clearly see a pronounced anisotropy, especially in the first row of the figure. More precisely, the shapes of the projected orbits in the first row are ellipses with the axes parallel to the diagonals of the corresponding coordinate systems. As we shall see, the anisotropy seen in the bottom row of Fig.~\ref{Fig-900K1StPairProjOrbits} will be mostly washed away when one averages over all existing pairs of $1^{\rm st}$-near neighbors. To further visualize the anisotropy, we projected the points onto the two diagonals and computed the histograms of the collapsed points, which are reported in the insets. The histograms reveal, in a more quantitative fashion, the anisotropy of the projected orbits. This type histograms will be discussed in detail in the following sections and will only be used here as a measure of this anisotropy. 
 
 \begin{figure}
\center
  \includegraphics[width=0.95\textwidth]{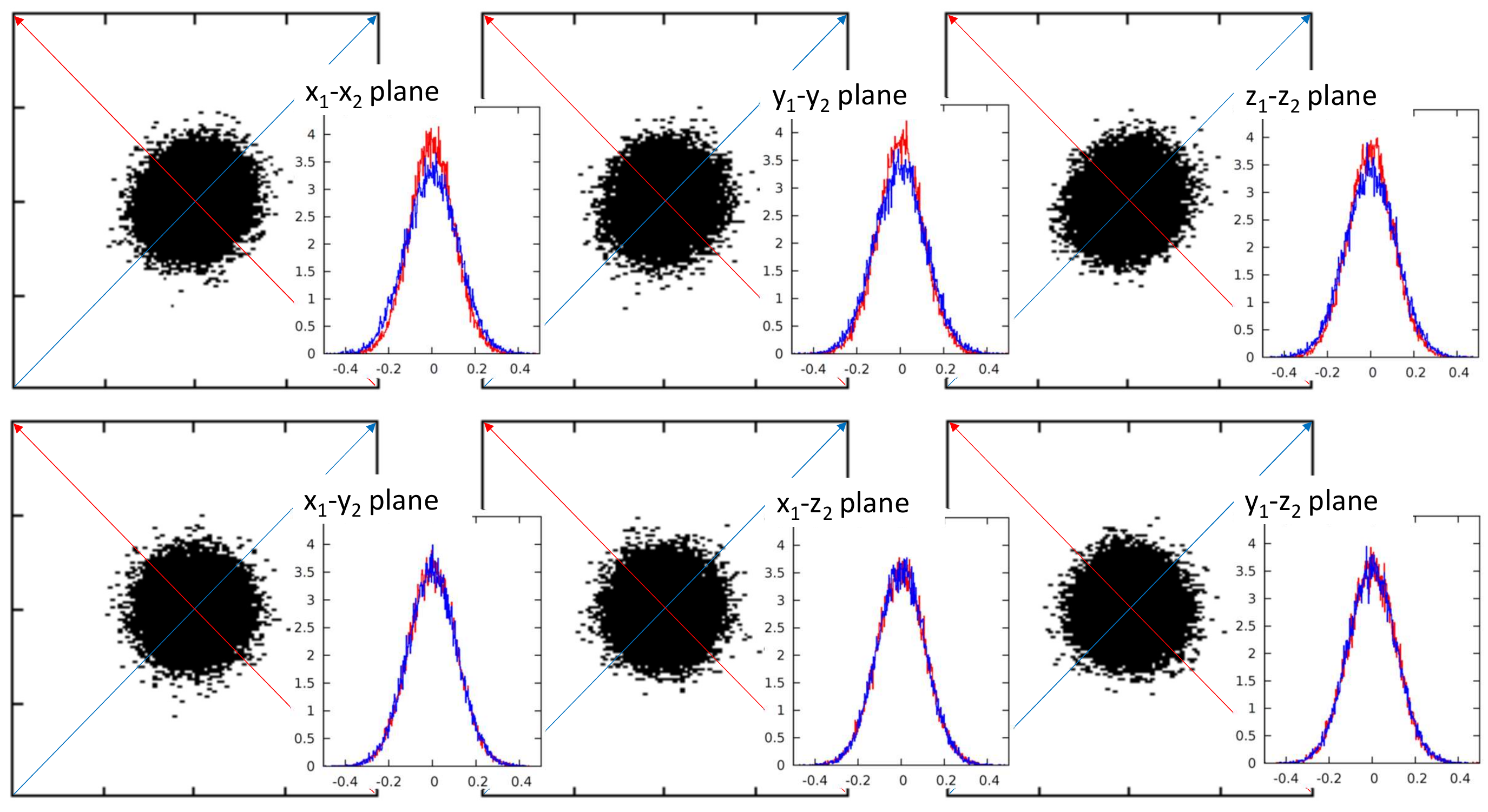}\\
  \caption{\small Same as Fig.~\ref{Fig-900K1StPairProjOrbits}, but for the $3^{\rm rd}$-near neighbors $\bm x_1=\bm x_{(0,0,0)}^{+1}$ and $\bm x_2=\bm x_{(1,1,-1)}^{-1}$.}
 \label{Fig-900K3RdPairProjOrbits}
\end{figure}

Figs.~\ref{Fig-900K2NdPairProjOrbits}, \ref{Fig-900K3RdPairProjOrbits} and \ref{Fig-900K4ThPairProjOrbits} report the results for pairs of $2^{\rm nd}$, $3^{\rm rd}$ and $4^{\rm th}$-near neighboring atoms, respectively. These projected orbits provide a representation of the sets $\Omega_n$, for $n=2,3,4$. Our data confirms that the correlation between the motions of the paired atoms diminishes with the rank $n$ of the pairs, which can be seen from the gradual fading of the anisotropy of the projected  orbits. In particular, the isotropy is almost restored for the pair of $4^{\rm th}$-near neighbors, and further analysis showed that the correlation between  $5^{\rm th}$ and higher near neighbors can be entirely ignored. We also want to point out that the anisotropy is absent in the bottom rows of these last three figures.

\subsection{Piecing Together the Configuration Space}

\begin{figure}
\center
  \includegraphics[width=\textwidth]{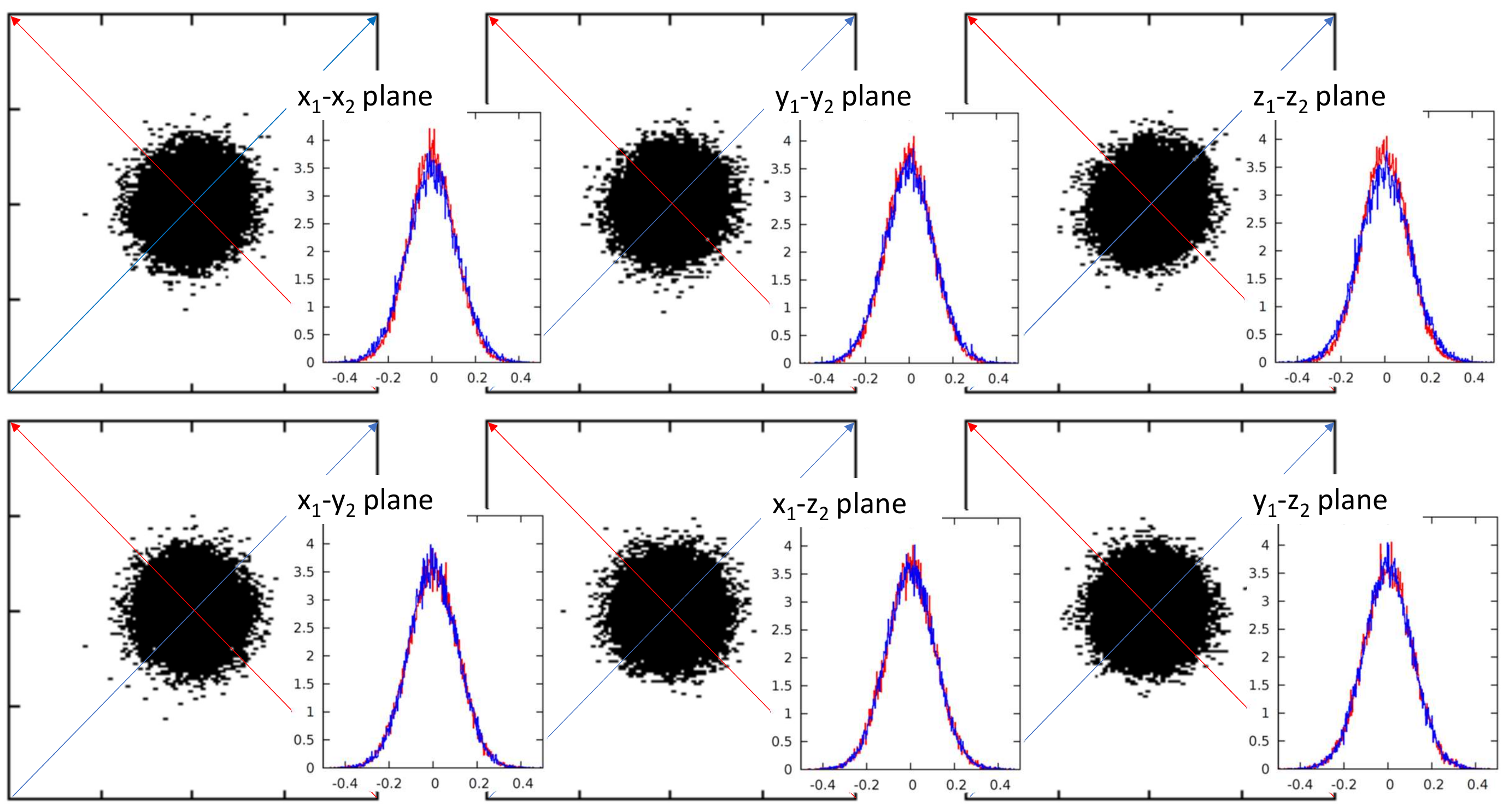}\\
  \caption{\small Same as Fig.~\ref{Fig-900K1StPairProjOrbits}, but for the $4^{\rm th}$-near neighbors $\bm x_1=\bm x_{(0,0,0)}^{-1}$ and $\bm x_2=\bm x_{(-1,1,1)}^{-1}$.}
 \label{Fig-900K4ThPairProjOrbits}
\end{figure}

We expect that the three-body correlations to be very small, if not completely absent, and, consequently, Figs.~\ref{Fig-900K1StPairProjOrbits}-\ref{Fig-900K4ThPairProjOrbits} contain all the information we need to piece together the configuration space of the crystal. In our opinion, this is a remarkable conclusion, because it means we can piece together the infinite set $\Omega$ from a finite amount of information, which can be directly extracted from the present CPMD simulations. This gives us hope that the configuration space of all crystalline materials can be accurately quantified in similar fashions. So, what is the shape of the configuration for the silicon crystal? Perhaps the most concise way to describe it is as a sculpted infinite dimensional sphere, such that:
\begin{itemize}

\item If $\Omega$ is projected on a plane made out of one coordinate of one atom and one coordinate of its first, second, third or forth neighboring atom, then the shape of the projection is a filled ellipse.  

\item The projection of $\Omega$ on any other 2-dimensional plane is a disk.

\end{itemize} 
These shapes will be quantified in the following sections.

\section{Analysis of the Gibbs Measure}
\label{Sec-HGibbsMeasure}

In this section we continue to make use of the projections, this time to quantify the Gibbs measure. In essence, our analysis resides on the fact that, if $f : \Omega \rightarrow \Gamma$ is a continuous function from $\Omega$ to some other topological space $\Gamma$, then $f$ generates an image of the Gibbs measure on the latter, defined by the so called push-forward measure:
\begin{equation}
(f_\ast \PM)(\Delta) = \PM\big ( f^{-1}(\Delta) \big ), \quad \Delta \subset \Gamma,
\end{equation}
or shortly $f_\ast \PM = \PM \circ f^{-1}$. For us, the role of $f$ will be played by various projections and, as we shall see, the push-forward measures can be quantified quite effectively from the numerical data. Then, as in the previous section, we will be able to piece together the Gibbs measure, this time from the histograms generated from the orbits.

\subsection{Temporal Histograms: Projecting on One Atom}
\label{Sec-THProjOneAtom}

As already mentioned above, the maps ${\rm P}_{\bm x}: \Omega \rightarrow \Omega_{0}$ induce the push-forward measures $\PM_{\bm x} = ({\rm P}_{\bm x})_* \PM$ over the set $\Omega_0$ studied in section~\ref{Sec-TemOrb1}. The goal of this section is to characterize, quantitatively, these measures. They can be interpreted as the measures which result from the Gibbs measure after all but $\omega_{\bm x}$ degrees of freedom are integrated out. Let us point out that, from symmetry and invariance considerations:
\begin{equation}
\PM_{\bm g \cdot \bm x} = \PM \circ {\rm P}_{\bm g \cdot \bm x}^{-1} = \PM \circ \tau_{\bm g^{-1}} \circ {\rm P}_{\bm x}^{-1} = \PM \circ {\rm P}_{\bm x}^{-1} = \PM_{\bm x},
\end{equation}
hence all these projected measures are identical with some measure $\PM_0$ on $\Omega_0$.

\vspace{0.2cm}

Let us introduce a useful notation, the indicator function $\chi_{A}$ of some generic set $A$, which takes value $1$ when evaluated inside $A$ and $0$ otherwise. From the ergodic assumption: 
\begin{equation}
\PM(A) = \lim_{T \rightarrow \infty} \frac{1}{T} \int_0^T {\rm d} t \, \chi_A\big ( \omega(t) \big ), \quad A \subset \Omega.
\end{equation}
Now, by definition:
\begin{equation}
\PM_{\bm x}(\Delta) = \PM\big ( ({\rm P}_{\bm x})^{-1}(\Delta) \big ), \quad \Delta \subset \Omega_0, 
\end{equation}
hence:
\begin{equation}
\PM_{\bm x}(\Delta) = \lim_{T \rightarrow \infty} \frac{1}{T} \int_0^T {\rm d} t \, \chi_{({\rm P}_{\bm x})^{-1}(\Delta)}\big ( \omega(t) \big ),
\end{equation}
or:
\begin{equation}\label{Eq-PPrinciple1}
\PM_{\bm x}(\Delta) = \lim_{T \rightarrow \infty} \frac{1}{T} \int_0^T {\rm d} t \, \chi_\Delta\big ( \omega_{\bm x}(t) \big ).
\end{equation}
This identity and variations of it represent our basic principle for mapping $\PM_{\bm x}$. 

\begin{figure}
\center
  \includegraphics[width=0.8\textwidth]{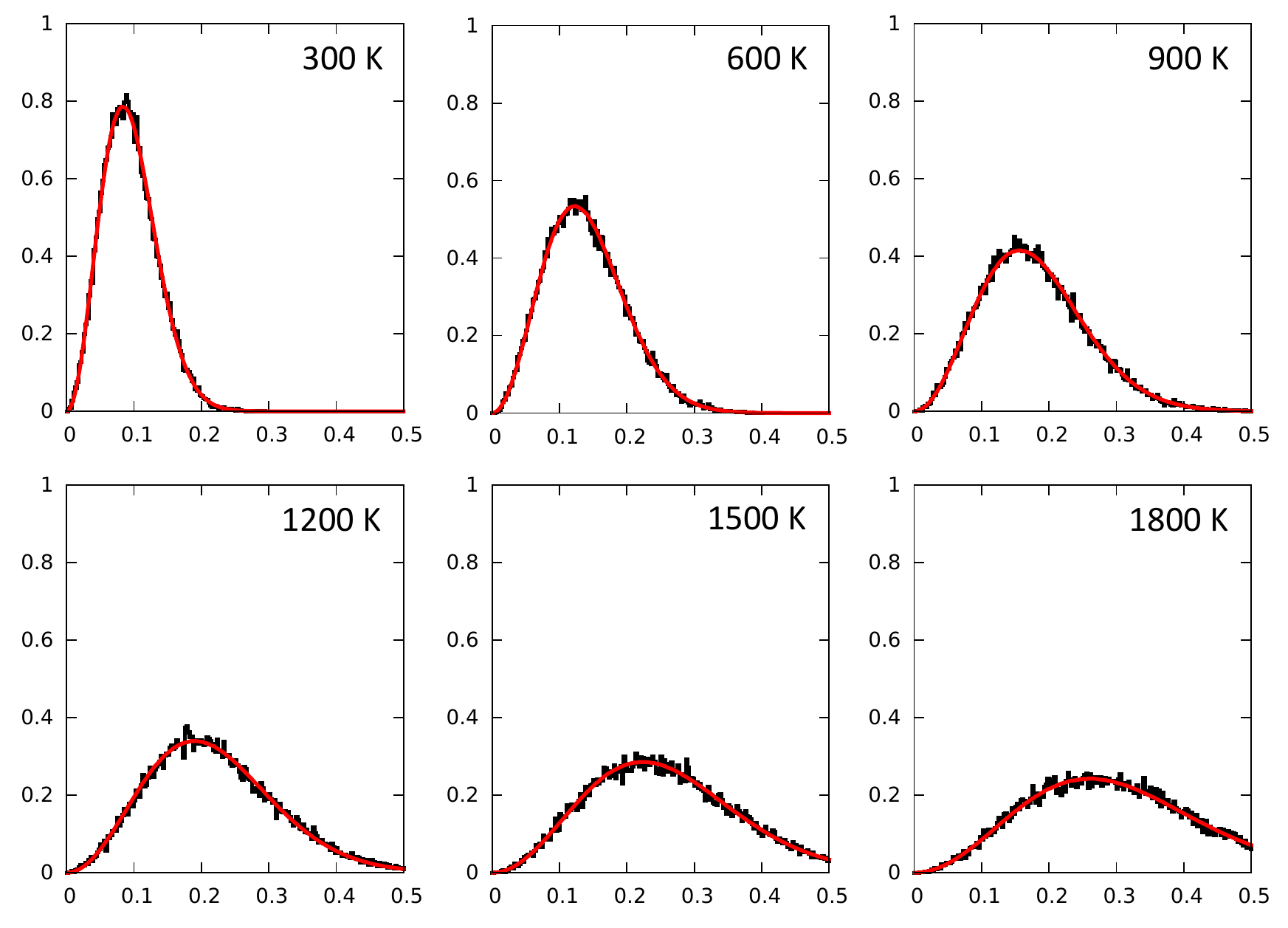}\\
  \caption{\small Histograms of the radial distributions of the atoms (relative to their average positions), as mapped from the temporal orbit of a single atom. The red curve represents an average over the histograms of all 1000 Si atoms; it is the most accurate representation of the radial distribution of the atoms. The histograms are mapped at different temperatures, ranging from $T=300\ K$ to $1800\ K$.}
 \label{OneTimeHistogram}
\end{figure}

\vspace{0.2cm}

In our computer simulations, the time interval is finite and the temporal orbits are discrete. If we denote the projected discrete orbits by:
\begin{equation}
\Oo_{\bm x} = \big \{\omega_{\bm x}(t_n), \ t_n=n \Delta t, \ n = 1, \ldots P \big \},
\end{equation}
where $P$ is the total number of recorded points in the time interval $[0,T]$, then:
\begin{equation}\label{Eq-PPrinciple2}
\widetilde \PM_{\bm x}(\Delta) = \frac{1}{T} \sum_{n=1}^P \chi_\Delta\big ( \omega_{\bm x}(t_n) \big ) \Delta t = \frac{|\Oo_{\bm x}\cap \Delta|}{P},
\end{equation}
provides an approximate representation of \eqref{Eq-PPrinciple1}, which will be the main tool for this section. Throughout, $|\cdot |$ represents the cardinal of a finite set. Furthermore:
\begin{equation}
\widetilde \PM_0 \big (\Delta \big )=\tfrac{1}{\Nn}\sum_{\bm x}\widetilde \PM_{\bm x}\big (\Delta \big ),
\end{equation} 
with $\Nn$ the number of atoms in our simulation, provides the most accurate numerical representation (given the available data) of the measure $\PM_0$.

\vspace{0.2cm}

\begin{figure}
\center
  \includegraphics[width=0.8\textwidth]{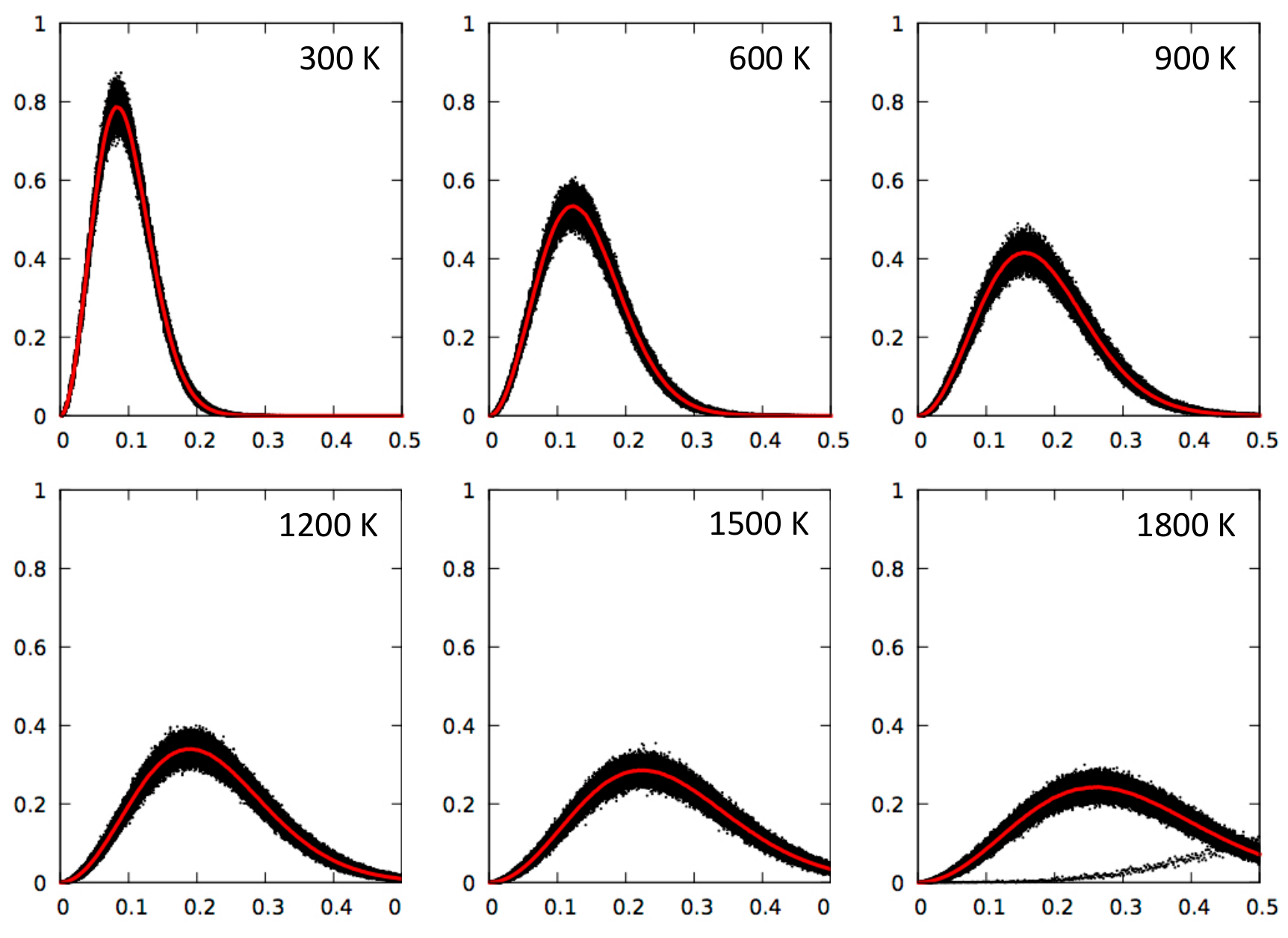}\\
  \caption{\small Superposition of all the histograms $\tfrac{1}{4\pi d}\widetilde \PM_{\bm x}\big (S_{r,d} \big )$ as constructed from the temporal orbits of the individual atoms. The red curves represent their averages. The histograms are mapped at different temperatures, ranging from $T=300\ K$ to $1800\ K$.}
 \label{TimeHistograms}
\end{figure}

Now, let $\Ss(r,d)$ be the spherical shell:
\begin{equation}
S_{r,d} = \big \{\vec r \in \RM^3 \, \big | \, r-\tfrac{d}{2} \leq |\vec r| \leq r + \tfrac{d}{2} \big \}.
\end{equation}
In Fig.~\ref{OneTimeHistogram}, we report a set of histograms representing $\tfrac{1}{4\pi d}\widetilde \PM_{\bm x_{\bm 0}^{-1}}\big (S_{r,d} \big )$ as a function of $r$, for various temperatures. The thickness of the shell was fixed at a small but, nevertheless, finite value, namely $d=0.0025$. This type of analysis provides information about the radial distributions and, as already implied by our notation, the histograms in Fig.~\ref{OneTimeHistogram} have been obtained from the orbit of a single atom (initially located at $\bm x_0^{-1}$). In Fig.~\ref{TimeHistograms}, we report the overlap all the histograms representing  $\tfrac{1}{4\pi d}\widetilde \PM_{\bm x}\big (S_{r,d} \big )$, corresponding to all individual atoms included in the simulations. If the orbits were infinitely long, then the histograms representing $\tfrac{1}{4\pi d}\widetilde \PM_{\bm x}\big (S_{r,d} \big )$ will be all the same. As such, the fluctuations that one sees in both Figs.~\ref{OneTimeHistogram} and \ref{TimeHistograms} are all due to the finite length of the temporal orbits. However, the principles of statistical mechanics assures us that we can compensate for this inherent computational reality by including enough atoms in the simulations and by taking an ensemble average. In Fig.~\ref{TimeHistograms}, the average histograms: 
\begin{equation}
\tfrac{1}{4\pi d}\widetilde \PM_0 \big (S_{r,d} \big )=\tfrac{1}{4\pi d \Nn}\sum_{\bm x}\widetilde \PM_{\bm x}\big (S_{r,d} \big ),
\end{equation} 
are represented by the red curves. These averages have also been superimposed over the data in Fig.~\ref{OneTimeHistogram}. If we accept the ergodicity hypothesis w.r.t. the crystal's space group, then these averages provide the best representations of $\tfrac{1}{4 \pi d} \PM_0\big (S_{r,d} \big )$ that can be obtained from the available data. With a number $\Nn=1000$ of Si atoms, the statistical error for the averages presented in Fig.~\ref{TimeHistograms} can be estimated to be less than 1\%. Of course, there are systematic errors due to finite-size effects \cite{Binder1996}, but for the size used in our simulation they are expected to be insignificant. Let us point out that, at $T=1800\ K$ in Fig.~\ref{TimeHistograms}, there are several orbits that leave the atomic site where the orbits originate, which, of course, is a sign of crystal melting.

\vspace{0.2cm}

\begin{figure}
\center
  \includegraphics[width=0.8\textwidth]{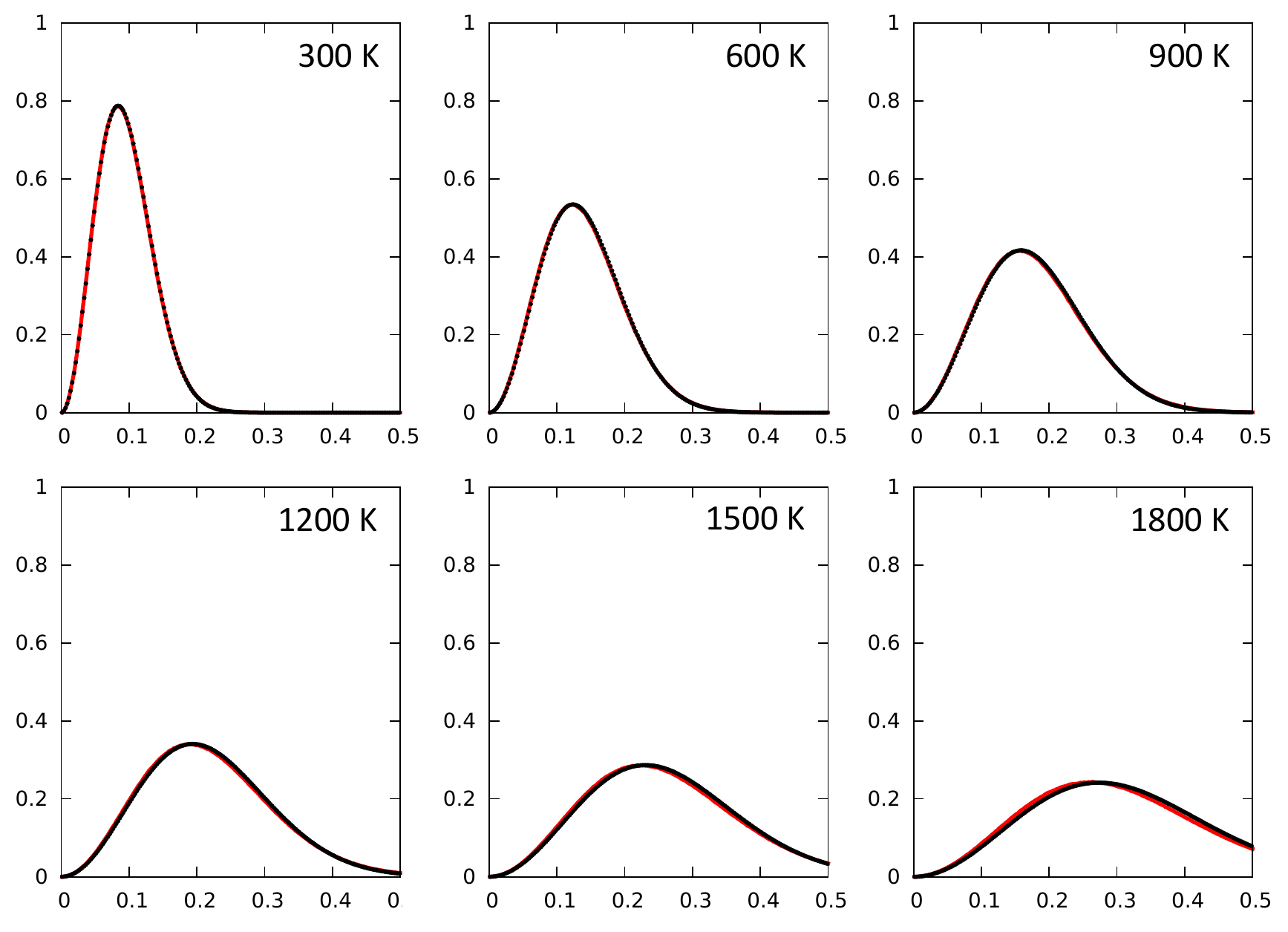}\\
  \caption{\small Comparison between the average histogram from Fig.~\ref{TimeHistograms} and the Guassian distribution $\frac{r^2}{(2 \pi \sigma_0^2)^\frac{3}{2}} e^{-\frac{r^2}{2 \sigma_0^2}} $, with the standard deviations computed by \eqref{Eq-StandardDeviation1}.}
 \label{FitAvgHistogram}
\end{figure}

By varying $d$, we found that the plots stabilize and become practically independent of $d$ at small values. This is an indication that the measures have densities, which we write below in spherical coordinates:
\begin{equation}
\PM_0 ({\rm d}^3 \bm r )=\rho_0(r,\sigma)r^2 {\rm d}r{\rm d}\sigma, \quad \sigma = (\theta,\phi), \quad {\rm d}\sigma = \sin \theta {\rm d}\theta {\rm d}\phi.
\end{equation} 
Note that:
\begin{equation}
\lim_{d \rightarrow 0} \tfrac{1}{4\pi d} \PM_0\big (S_{r,d} \big ) = \tfrac{r^2}{4\pi}\int \rho_0(r,\sigma) {\rm d}\sigma = r^2 \big \langle \rho_0(r,\sigma) \big \rangle_\sigma,
\end{equation}
which explains our factorization constant $\tfrac{1}{4\pi d}$. Above, $\langle \cdot \rangle_\sigma$ denotes the average over the solid angle $\sigma$. From the data reported in Fig.~\ref{TimeHistograms}, we found that the probability measures $\PM_0$ is extremely well represented by a product of normal distributions:
\begin{equation}
 \PM_0({\rm d}^3 \vec r) \approx \prod_{i=1}^3 \tfrac{1}{\sqrt{2\pi \sigma_0^2}}\exp\Big (-\tfrac{x_i^2}{2 \sigma_0^2} \Big ) {\rm d}x_i,
 \end{equation}
which in spherical coordinates reads:
\begin{equation}
\prod_{i=1}^3 \tfrac{1}{\sqrt{2\pi \sigma_0^2}}\exp\Big (-\tfrac{x_i^2}{2 \sigma_0^2} \Big ) {\rm d}x_i = \tfrac{1}{(2\pi \sigma_0^2)^\frac{3}{2}} \exp\Big (-\tfrac{r^2}{2 \sigma_0^2} \Big ) r^2 {\rm  d}r{\rm d} \sigma.
\end{equation}
We compute the standard deviation from the average histogram in Fig.~\ref{TimeHistograms}, as:
 \begin{equation}\label{Eq-StandardDeviation1}
3 \sigma_0^2 = \langle r^2 \rangle =  d \sum_{n \in \NM} r_n^2 \, \widetilde \PM_0\big (S_{r_n,d} \big ) , \quad r_n = (n+\tfrac{1}{2})d,
\end{equation}
and, with this choice, the overlap between $(2\pi \sigma_0^2)^{-\frac{3}{2}} r^2 e^{-\frac{r^2}{2 \sigma_0^2}} $ and $\tfrac{1}{4\pi d}\widetilde \PM_0 \big (S_{r,d} \big )$ is almost perfect, as shown in Fig.~\ref{FitAvgHistogram}, with perhaps the exception of the highest temperature. This is one of the fundamental findings of our work since it shows that the radial distribution $\big \langle \rho_0(r,\sigma) \big \rangle_\sigma$ can be quantified almost perfectly using a single parameter. 

\vspace{0.2cm}

\begin{figure}
\center
  \includegraphics[width=0.6\textwidth]{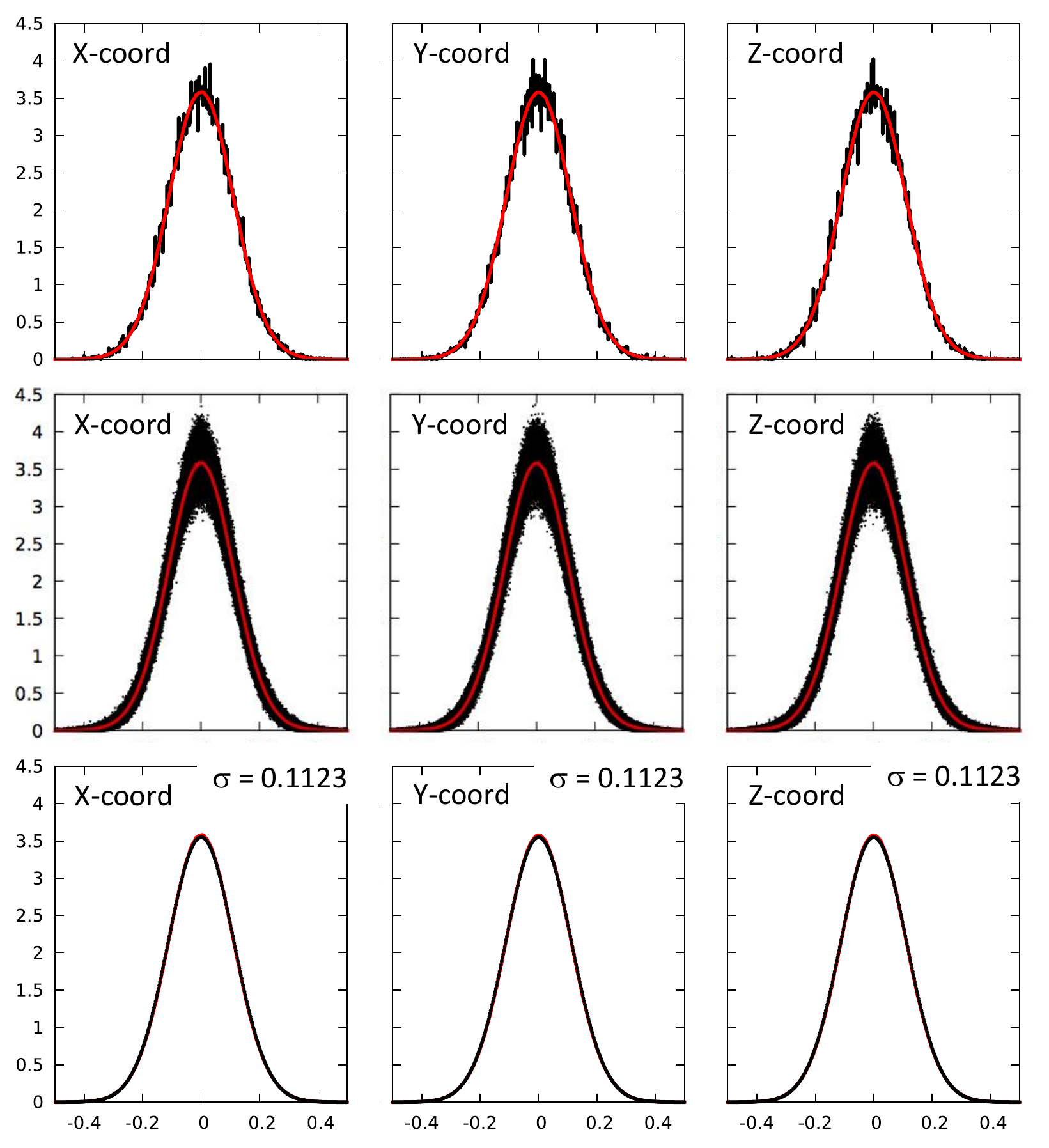}\\
  \caption{\small Resolution of the histograms from Fig.~\ref{TimeHistograms} along different space directions. The histograms in the first row were extracted from the orbit of a single atom. The second row shows a superposition of such histograms when we ran over all atoms in the simulation, together with the corresponding averages (red curve). The third row shows a comparison between the averages and a Gaussian distribution, whose standard deviation is also displayed. Note the almost perfect overlap.}
 \label{XYZTimeHistograms}
\end{figure}

To probe the angular dependence of the distribution $\rho_0(r,\sigma)$, we resolve the histograms in each space direction and the results are reported in Fig.~\ref{XYZTimeHistograms} for $T=900 \ K$. The first row of this figure corresponds to data extracted from the orbit of a single atom, more precisely, it plots $\frac{1}{d}\frac{|\Oo{\bm x_{\bm 0}^{-1}}\cap \Delta_{u}(s)|}{P}$ as a function of $s$, where:
\begin{equation}
\Delta_u(s) = \Big \{\bm r \in  \Omega_0, \ s-\tfrac{d}{2} \leq u \leq s + \tfrac{d}{2} \Big \}, \quad u=x,y,z.
\end{equation}
In the second row we show a superposition of such histograms when we run over all atoms in the simulations, together with the average histograms shown by the red curves. In the third row we show a comparison between these averages and the normal distribution $\frac{1}{\sqrt{2\pi \sigma_0^2}}\exp\Big (-\tfrac{s^2}{2 \sigma_0^2}\Big )$ corresponding to the same standard deviation as in Fig.~\ref{FitAvgHistogram} and whose numerical value is shown in Fig.~\ref{XYZTimeHistograms}. The overlap between the two are almost perfect, which a quantitative confirmation of our previous observation, drawn from Figs.~\ref{TempDepOrbits} and \ref{Fig-OneAtomOrbit}, that there is practically no angular dependence in these distributions. Similar conclusion applies for all the other simulated temperatures.

\subsection{Temporal Histograms: Projecting on Two Atom}

Let us consider the projection on the coordinates of two $n^{\rm th}$-near neighboring atoms:
\begin{equation}
{\rm P}_{\bm x_1,\bm x_2} \omega = \big (\omega_{\bm x_1},\omega_{\bm x_2} \big ) \in {\rm P}_{\bm x_1,\bm x_2}( \Omega) \subset \Omega_0 \times \Omega_0, \quad ({\bm x_1,\bm x_2}) \in \Pp_n.
\end{equation}
Recall \eqref{Eq-Omega1}, which says that all sets ${\rm P}_{\bm x_1,\bm x_2}( \Omega)$ are identical. This unique set was denoted by $\Omega_n$ in the previous section. The projection then provides the push-forward measure $\PM_{\bm x_1,\bm x_2} = \big ( {\rm P}_{\bm x_1,\bm x_2} \big )_\ast \PM$ over $\Omega_n$, for any $({\bm x_1,\bm x_2}) \in \Pp_n$, which can be interpreted as the measure resulting from the Gibbs measure after all but $(\omega_{\bm x_1},\omega_{\bm x_2})$ degrees of freedom are integrated out. Again, from symmetry and invariance considerations:
\begin{equation}
\PM_{\bm g \cdot \bm x_1,\bm g \cdot \bm x_2} = \PM \circ {\rm P}_{\bm g \cdot \bm x_1,\bm g \cdot \bm x_2}^{-1} = \PM \circ \tau_{\bm g^{-1}} \circ {\rm P}_{\bm x_1,\bm x_2}^{-1} = \PM \circ {\rm P}_{\bm x_1,\bm x_2}^{-1} = {\rm P}_{\bm x_1,\bm x_2},
\end{equation}
and, since all the $n^{\rm th}$-near neighboring pairs of atoms can be transformed into each other by space group transformations, all these projected measures are identical with some measure $\PM_n$ on $\Omega_n$. Furthermore, from similar considerations as in section~\ref{Sec-THProjOneAtom}:
\begin{equation}\label{Eq-PPrinciple3}
\PM_{\bm x_1,\bm x_2}(\Delta) = \lim_{T \rightarrow \infty} \frac{1}{T} \int_0^T {\rm d} t \, \chi_\Delta\big ( \omega_{\bm x_1}(t), \omega_{\bm x_2}(t) \big ), \quad  \Delta \subset \Omega_n.
\end{equation} 
If we denote the new class of projected discrete orbits by:
\begin{equation}
\Oo_{\bm x_1,\bm x_2} = \Big \{ \big (\omega_{\bm x_1}(t_j), \omega_{\bm x_2}(t_j) \big ), \ t_j=j\,  \Delta t, \ j = 1, \ldots P \Big \},
\end{equation}
then:
\begin{equation}\label{Eq-PPrinciple4}
\widetilde \PM_{\bm x_1,\bm x_2}(\Delta) = \frac{1}{T} \sum_{j=1}^P \chi_\Delta\big ( \omega_{\bm x_1}(t_j),  \omega_{\bm x_2}(t_j) \big ) \Delta t = \frac{|\Oo_{\bm x_1,\bm x_2}\cap \Delta|}{P},
\end{equation}
provides an approximate representation of \eqref{Eq-PPrinciple3}, which will be the main tool for this section. Furthermore, if $\widetilde \Pp_n$ denotes the set of all $n^{\rm th}$-near neighboring pairs of atoms in our simulations, then:
\begin{equation}
\widetilde \PM_n(\Delta)=\frac{1}{|\widetilde \Pp_n|}\sum_{(\bm x_1,\bm x_2) \in \widetilde \Pp_n} \widetilde \PM_{\bm x_1,\bm x_2}(\Delta), \quad \Delta \in \Omega_n,
\end{equation}
provides the best numerical representation (from the available data) of the measures $\PM_n$. 

\begin{figure}
\center
  \includegraphics[width=1\textwidth,angle=90]{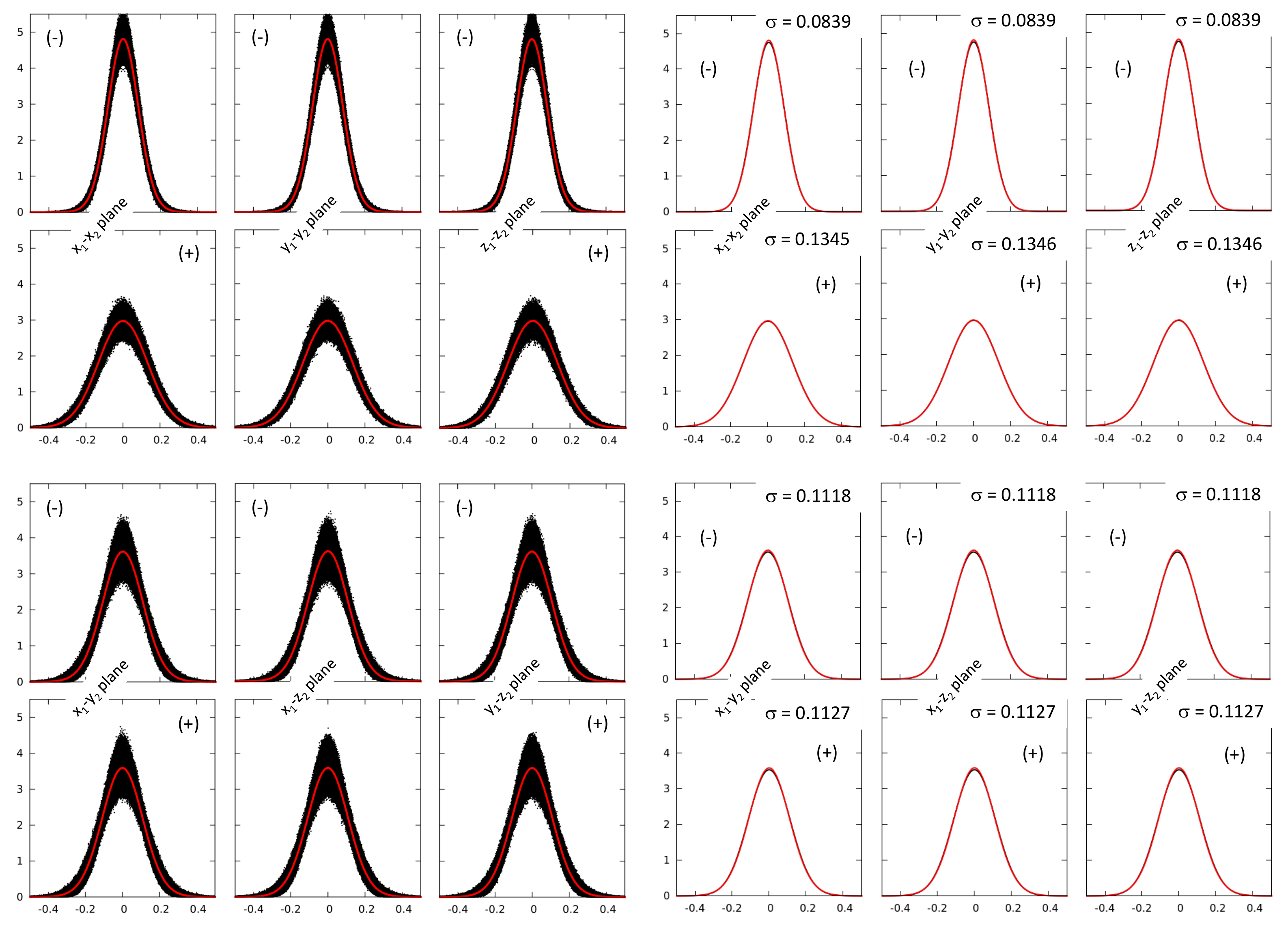}\\
  \caption{\small Histograms representing $\tfrac{1}{d} \widetilde \PM_{\bm x_1,\bm x_2}\big ( \Delta^{\pm}_{u_1,u_2}(s) \big )$ as a function of the parameter $s$, for $(\bm x_1,\bm x_2) \in \Pp_1$ and combinations of $u_1=x_1,x_2,x_3$ and $u_2=x_2,y_2,z_2$, respectively. The sets $\Delta^{\pm}_{u_1,u_2}(s)$ are defined in \eqref{Eq-DeltaS} and $d$ was fixed at $0.0025$. The panels on the left show the overlap of the histograms corresponding to all $1^{\rm st}$-near neighboring pairs of atoms in our numerical simulation (black dots), together with the averages of these histograms (red curves). The panels on the right report a comparison between the averaged histograms (red curves) and the normal distributions (black curves) $\frac{1}{\sqrt{2\pi \sigma^2}} e^{-\frac{s^2}{2\sigma^2}}$, with $\sigma$ reported in each panel.}
 \label{Fig-1StPairCorrelations}
\end{figure}

\vspace{0.2cm}

Let us point out that it is the measures $\PM_{\bm x_1,\bm x_2}$ that quantify the distributions rendered in Figs.~\ref{Fig-900K1StPairProjOrbits}-\ref{Fig-900K4ThPairProjOrbits}. Therein, the distributions were resolved on various coordinate planes $(u_1,u_2)$, with $u_1=x_1,y_1,z_1$ and $u_2=x_2,y_2,z_2$,  and it is found that the distributions display mirror symmetries w.r.t. the diagonals of those planes. Hence, it is more appropriate to use these diagonals as coordinates and make a change of variables:
\begin{equation}
(u_1,u_2) \rightarrow \Big ( \tfrac{1}{\sqrt{2}}(u_1 -u_2),\tfrac{1}{\sqrt{2}}(u_1+u_2) \Big ).
\end{equation} 
As such, we will be dealing with histograms representing: 
\begin{equation}
\tfrac{1}{d} \widetilde \PM_{\bm x_1,\bm x_2}\big ( \Delta^{\pm}_{u_1,u_2}(s) \big ) = \frac{1}{d}\frac{|\Oo_{\bm x_1,\bm x_2}\cap \Delta^{\pm}_{u_1,u_2}(s)|}{P}
\end{equation} as a function of $s$, where:
\begin{equation}\label{Eq-DeltaS}
 \Delta_{u_1,u_2}^{\pm}(s) = \Big \{(\bm r_1, \bm r_2) \in  \RM^6 , \ s-\tfrac{d}{2} \leq \tfrac{1}{\sqrt{2}} \big ( u_1 \pm u_2 \big ) \leq s + \tfrac{d}{2} \Big \}.
\end{equation}
As such, for each pair of atoms and pair of space directions $(u_1,u_2)$, there will be two plots labeled by (+/-) and corresponding to the first/second diagonals in Figs.~\ref{Fig-900K1StPairProjOrbits}-\ref{Fig-900K4ThPairProjOrbits}. Some of these histograms have been already plotted in the insets of these figures. Again, after gradually decreasing the value of $d$, we see that $\tfrac{1}{d} \widetilde \PM_{\bm x_1,\bm x_2}\big ( \Delta^{\pm}_{u_1,u_2}(s) \big )$ becomes independent of this parameter, hence the measures have densities and we can write:
\begin{equation}
\PM_{\bm x_1,\bm x_2}({\rm d}^3 \bm r_1 \times {\rm d}^3 \bm r_2) = \rho_n(\bm r_1,\bm r_2) {\rm d}^3 \bm r_1 \, {\rm d}^3 \bm r_2, \quad ({\bm x_1,\bm x_2}) \in \Pp_n.
\end{equation}
Now, by construction, if we integrate out $\bm r_2$, we should recover the measure studied in section~\ref{Sec-THProjOneAtom}. As such, there is a direct relation between the two densities:
\begin{equation}\label{Eq-Check1}
\int {\rm d}^3 \bm r_2 \; \rho_n(\bm r_1,\bm r_2) = \rho_0(\bm r_1),
\end{equation} 
which provides an important check point for our analysis.

\begin{figure}
\center
  \includegraphics[width=1\textwidth,angle=90]{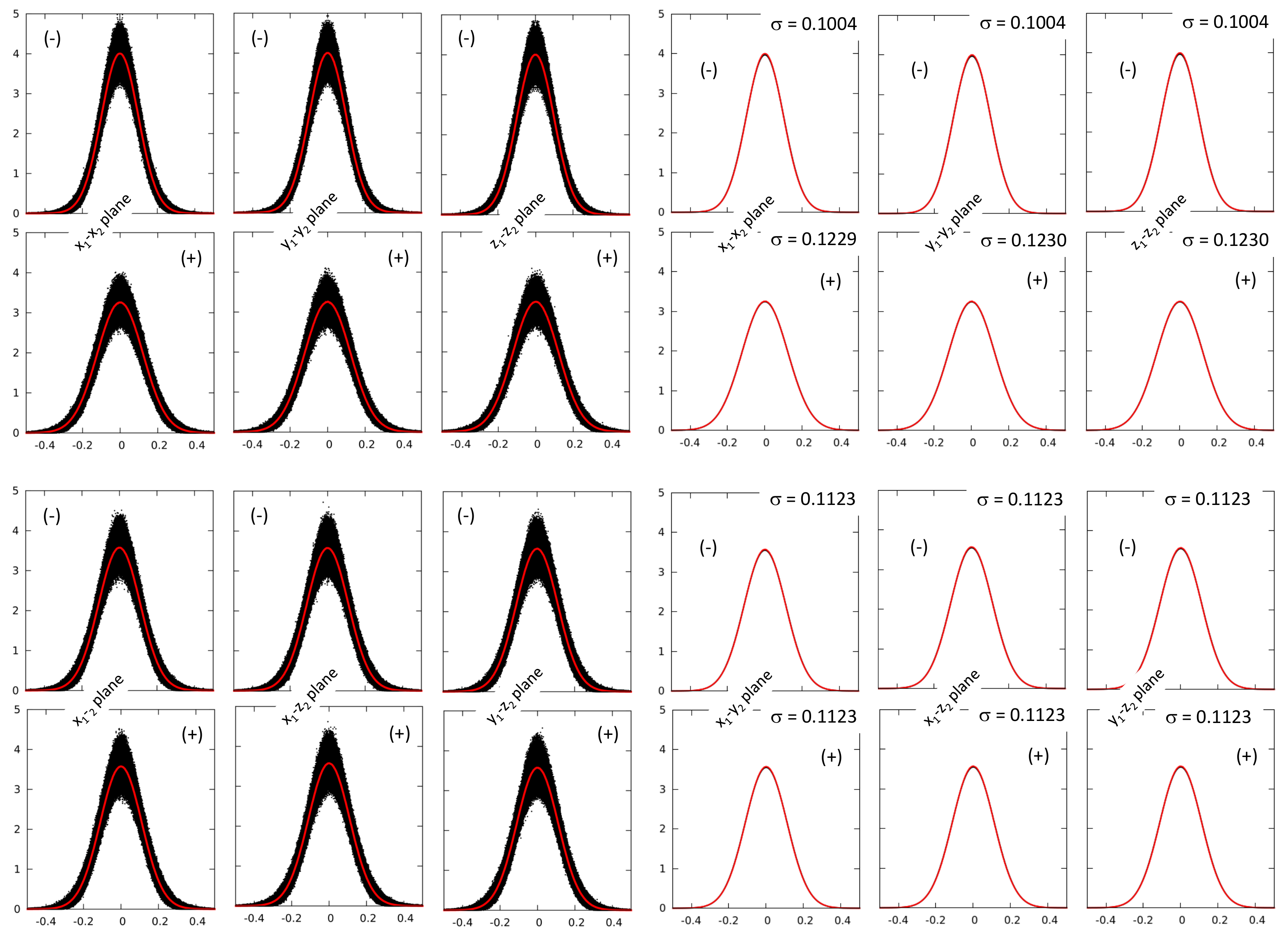}\\
  \caption{\small Same as Fig.~\ref{Fig-1StPairCorrelations}, but for pairs of $2^{\rm nd}$-near neighboring atoms.}
 \label{Fig-2NdPairCorrelations}
\end{figure}

\vspace{0.2cm}

The histograms for the $1^{\rm st}$-near neighboring atoms are reported in Fig.~\ref{Fig-1StPairCorrelations}. The left group of panels show an overlap of $\tfrac{1}{d} \widetilde \PM_{\bm x_1,\bm x_2}\big ( \Delta^{\pm}_{u_1,u_2}(s) \big )$ for all $1^{\rm st}$-near neighboring pairs of atoms from our simulations. The fluctuations seen in the left group of panels of Fig.~\ref{Fig-1StPairCorrelations} are due to the finite length of the temporal orbits. The same panels also report the ensemble average of the histograms:
\begin{equation}
\frac{1}{|\widetilde \Pp_1|}\sum_{(\bm x_1,\bm x_2)\in \widetilde \Pp_1} \tfrac{1}{d} \widetilde \PM_{\bm x_1,\bm x_2}\big ( \Delta^{\pm}_{u_1,u_2}(s) \big ), 
\end{equation}
which provides the best representation of $\rho_1$ that can be extracted from our numerical data. As in section~\ref{Sec-THProjOneAtom}, we find again that this distribution is extremely well approximated by a normal distribution. To quantify this statement, we have computed the standard deviations of the averaged histograms, generated normal distributions with the same standard distributions and then compared the two sets of distributions in the right group of panels in Fig.~\ref{Fig-1StPairCorrelations}. As one can see, the overlap is almost perfect.  

\begin{figure}
\center
  \includegraphics[width=1\textwidth,angle=90]{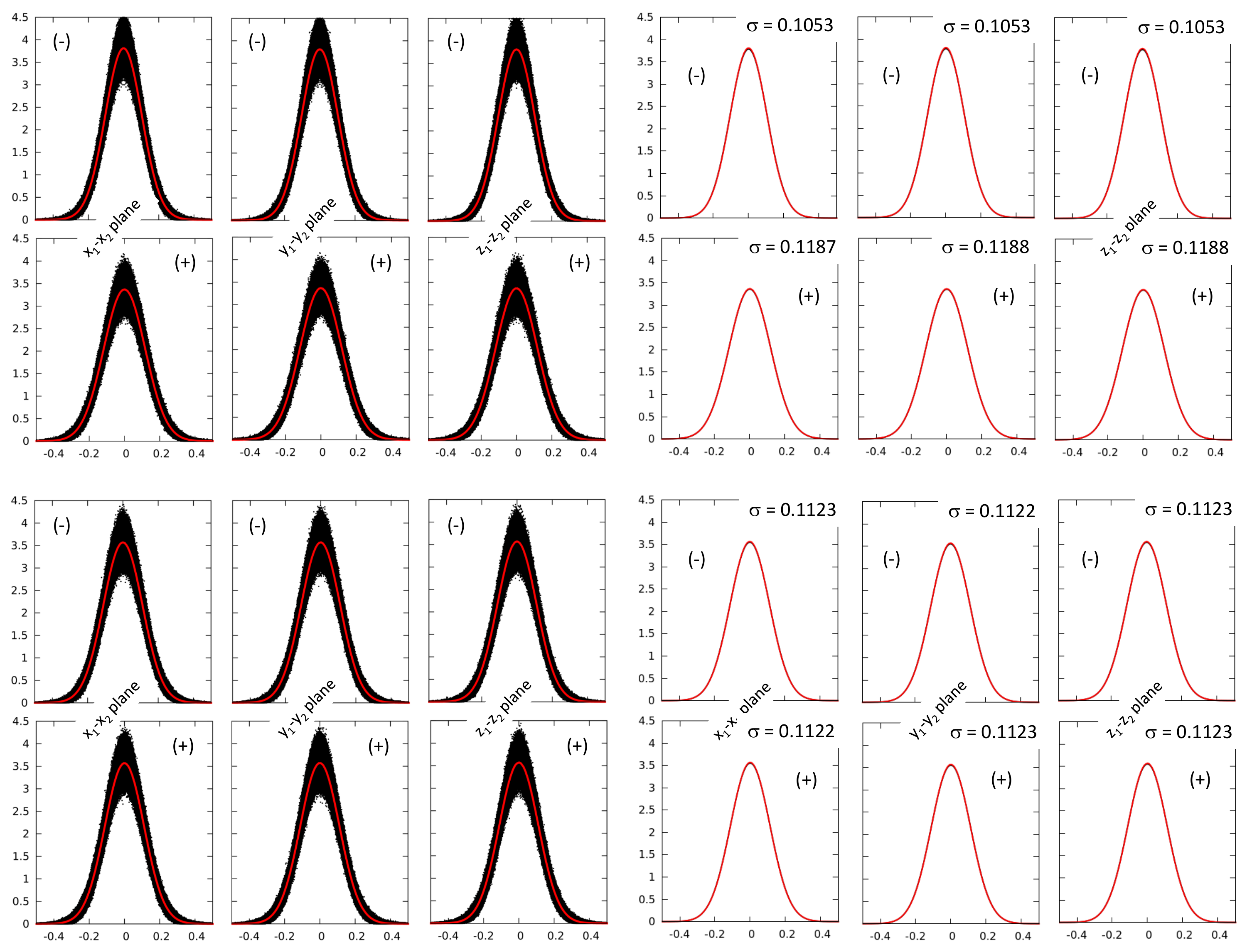}\\
  \caption{\small Same as Fig.~\ref{Fig-1StPairCorrelations}, but for pairs of $3^{\rm rd}$-near neighboring atoms.}
 \label{Fig-3RdPairCorrelations}
\end{figure}

\vspace{0.2cm}

Based on the data from Fig.~\ref{Fig-1StPairCorrelations}, we make the ansatz:
\begin{equation}\label{Eq-Ansatz1}
\rho_1(\bm r_1,\bm r_2)= \tfrac{1}{(4\pi^2 \alpha_1^2\beta_1^2)^{3/2}} \exp \Big ( -\tfrac{1}{2\alpha_1^2} \Big ( \tfrac{|\bm r_1-\bm r_2|}{\sqrt{2}}\Big )^2 - \tfrac{1}{2\beta_1^2} \Big ( \tfrac{|\bm r_1+\bm r_2|}{\sqrt{2}}\Big )^2 \Big ).
\end{equation} 
It is instructive to expand the exponential as:
\begin{equation}
\exp \Big ( -\Big (\tfrac{1}{4\alpha_1^2} + \tfrac{1}{4\beta_1^2}\Big ) \big (|\bm r_1|^2 + |\bm r_2|^2 \big ) \Big ) \exp \Big ( \Big (\tfrac{1}{2\alpha_1^2} - \tfrac{1}{2\beta_1^2}\Big ) \bm r_1 \cdot \bm r_2 \Big ),
\end{equation}
in which case we can see explicitly that, if $\alpha_1=\beta_1$, then the distribution reduces to an isotropic normal distribution over the 6-dimensional space, with standard deviation $\sigma = \alpha_1=\beta_1$. In this case, given the constraint \eqref{Eq-Check1}, the distribution factors as $\rho_1(\bm r_1,\bm r_2) = \rho_0(\bm r_1) \rho_0(\bm r_2)$ and, as such, the motions of the two atoms are completely un-correlated. If $\alpha_1 \neq \beta_1$, then the motions or the two atoms become correlated and the distributions become anisotropic, as already witnessed in Figs.~\ref{Fig-900K1StPairProjOrbits}-\ref{Fig-900K4ThPairProjOrbits}. Explicit computations of various correlation functions will be reported in next sections.

\begin{figure}
\center
  \includegraphics[width=1\textwidth,angle=90]{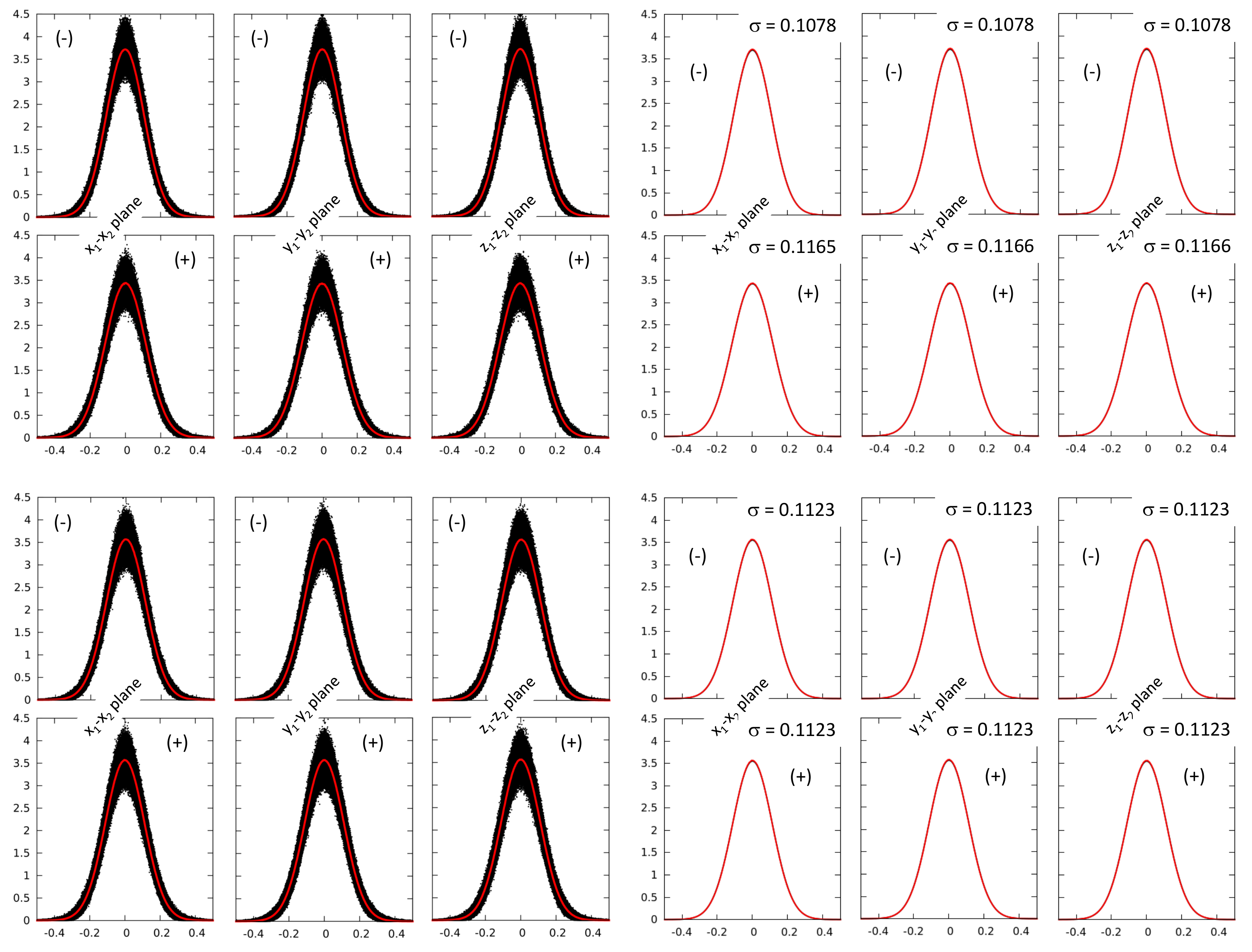}\\
  \caption{\small Same as Fig.~\ref{Fig-1StPairCorrelations}, but for pairs of $4^{\rm th}$-near neighboring atoms.}
 \label{Fig-4ThPairCorrelations}
\end{figure}

\vspace{0.2cm}

\begin{figure}
\center
  \includegraphics[width=0.7\textwidth,angle=0]{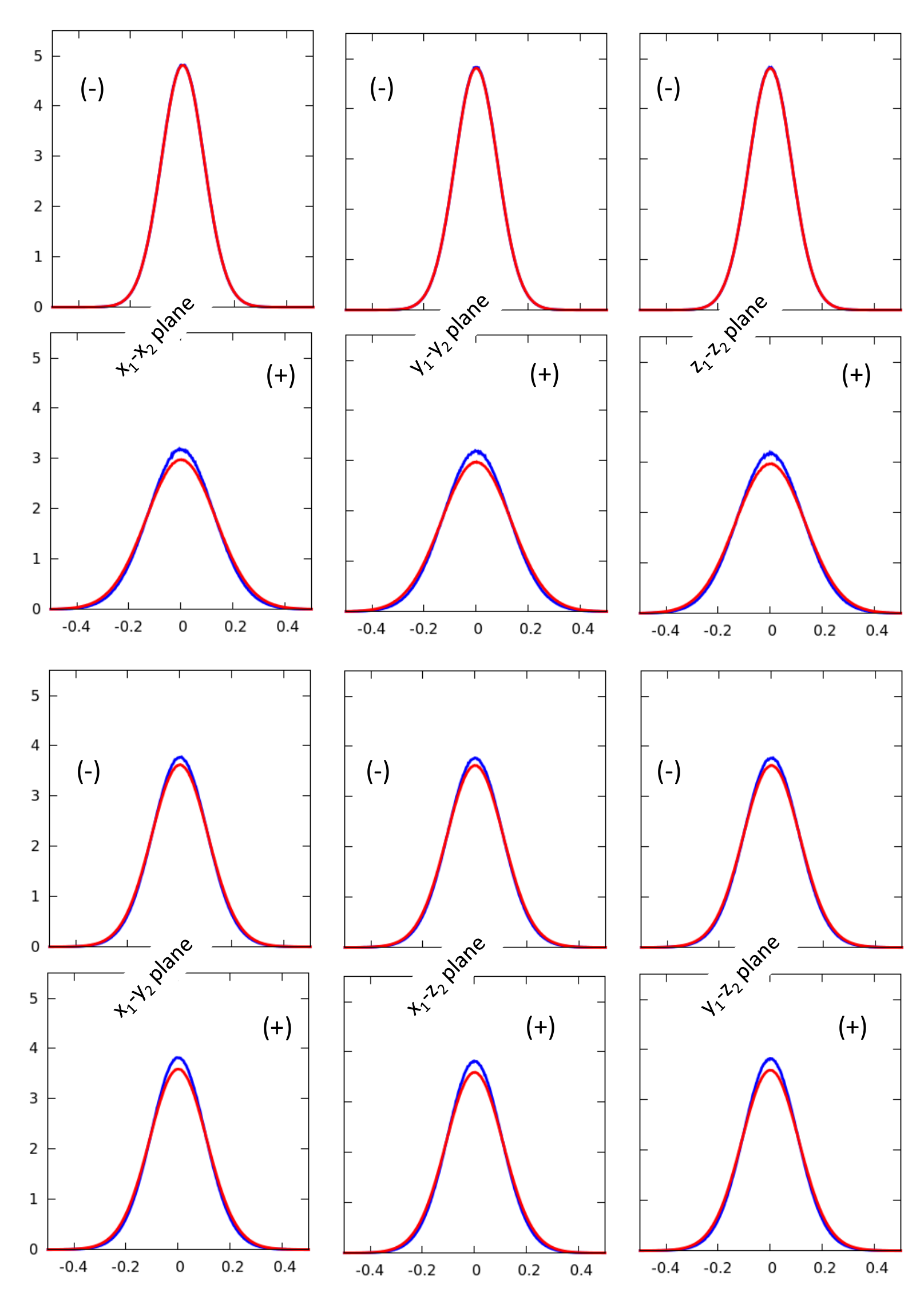}\\
  \caption{\small The averaged histograms from Fig.~\ref{Fig-1StPairCorrelations} for the first-nearest neighbor pairs of the 1000 atoms crystal are replotted (red curves) and compared with the same data computed from the CPMD simulations for a crystal containing only 216 atoms (blue curves).}
 \label{Fig-1StPairSizeComp}
\end{figure}

To make connection between the proposed distribution and the numerical data from Fig.~\ref{Fig-1StPairCorrelations}, we integrate all degrees of freedom except for a pair of $(u_1,u_2)$ coordinates. When this pair is $(x_1,x_2)$, the result of the integration is:
\begin{equation}
\tfrac{1}{2\pi \alpha_1 \beta_1} \exp \Big ( -\tfrac{1}{2\alpha_1^2} \Big ( \tfrac{|x_1-x_2|}{\sqrt{2}}\Big )^2 - \tfrac{1}{2\beta_1^2} \Big ( \tfrac{|x_1 + x_2|}{\sqrt{2}}\Big )^2 \Big ),
\end{equation}
and we can read off both $\alpha_1$ and $\beta_1$ from the distributions in the panels marked with the label $(x_1-x_2)$ in Fig.~\ref{Fig-1StPairCorrelations}. The conclusion is that $\alpha_1=0.0839$ and $\beta_1=0.1345$. The units for these coefficients are Angstroms. We could have chosen the $(y_1-y_2)$ or $(z_1-z_2)$ instead of $(x_1-x_2)$, but the conclusion would be the same, given the values of $\sigma$ reported in Fig.~\ref{Fig-1StPairCorrelations}. We still need to verify that these values are consistent with the rest of the data in Fig.~\ref{Fig-1StPairCorrelations}. Let us consider first the test where we integrate all degrees of freedom except of $(x_1,y_2)$. The result of such an integration is:
\begin{equation}
\exp \Big ( - \tfrac{x_1^2 + y_2^2}{\alpha_1^2 + \beta_1^2} \Big ) = \exp \Big ( -\tfrac{\big ((x_1-y_2)/\sqrt{2}\big )^2 + \big ((x_1+y_2)/\sqrt{2}\big )^2}{\alpha_1^2 + \beta_1^2} \Big ),
\end{equation}
hence we need to compare $\sqrt{ \tfrac{\alpha_1^2 + \beta_1^2}{2}}=0.1120$ with the standard variations reported in the panels labeled as $(x_1-y_2)$ in Fig.~\ref{Fig-1StPairCorrelations}. There are two panels and two distributions labeled that way, but the standard deviations reported in those panels are very close. We will take the view here that they are practically identical normal distributions with the average standard deviation, $\sigma=\tfrac{1}{2}(0.1118+0.1127)=0.1123$.  The first three digits of this value are identical to those of $\sqrt{ \tfrac{\alpha_1^2 + \beta_1^2}{2}}$. Furthermore, since the standard deviations reported in the other panels at the bottom of Fig.~\ref{Fig-1StPairCorrelations} are the same, we now can see that the ansatz \eqref{Eq-Ansatz1} is consistent with the entire data reported in Fig.~\ref{Fig-1StPairCorrelations}. Lastly, let us verify \eqref{Eq-Check1}. If we integrate all $\bm r_2$ degrees of freedom, the result is:
\begin{equation}
\tfrac{1}{\sqrt{\pi (\alpha_1^2 + \beta_1^2)}} \exp \Big ( - \tfrac{|\bm r_1|^2}{\alpha_1^2 + \beta_1^2} \Big ),
\end{equation}
hence we need to compare $\sqrt{ \tfrac{\alpha_1^2 + \beta_2^2}{2}}=0.1120$ with $\sigma_0=0.1123$. As one can see, the matching is again quite remarkable.

\vspace{0.2cm}

The data and the analysis for pairs of $2^{\rm nd}$, $3^{\rm rd}$ and $4^{\rm th}$-near neighboring atoms are reported in Figs.~\ref{Fig-2NdPairCorrelations}, \ref{Fig-3RdPairCorrelations} and \ref{Fig-4ThPairCorrelations}, respectively. For these cases, we find again that the ansatz:
\begin{equation}\label{Eq-Ansatz2}
\rho_n(\bm r_1,\bm r_2)= \tfrac{1}{(4\pi^2 \alpha_n^2\beta_n^2)^{3/2}} \exp \Big ( -\tfrac{1}{2\alpha_n^2} \Big ( \tfrac{|\bm r_1-\bm r_2|}{\sqrt{2}}\Big )^2 - \tfrac{1}{2\beta_n^2} \Big ( \tfrac{|\bm r_1+\bm r_2|}{\sqrt{2}}\Big )^2 \Big ) 
\end{equation} 
characterizes the numerical data with amazing precision. The numerical values of the coefficients are given below:
\begin{itemize}
\item $2^{\rm nd}$-near neighbors: $\alpha_2 = 0.1004$, $\beta_2=0.1230$, $\sqrt{\frac{\alpha_2^2 + \beta_2^2}{2}} = 0.1123$;
\item $3^{\rm rd}$-near neighbors: $\alpha_3 = 0.1053$, $\beta_3=0.1188$, $\sqrt{\frac{\alpha_3^2 + \beta_3^2}{2}} = 0.1123$;
\item $4^{\rm th}$-near neighbors: $\alpha_4 = 0.1078$, $\beta_4=0.1166$, $\sqrt{\frac{\alpha_4^2 + \beta_4^2}{2}} = 0.1123$.
\end{itemize}
Examining these values, one can see that all the tests hold true up to four digits of precision. Let us point out the convergence of $\alpha_n$ and $\beta_n$ towards $\sigma_0 = 0.1123$ as the rank $n$ of the pairs is increased.

\vspace{0.2cm}

Lastly, we address the convergence of the results with the size of the crystal. For this, we repeated the entire analysis for a Si crystal containing only 216 atoms, that is, $3 \times 3 \times 3$ unit cells. Except for the size of the crystal, all parameters of the simulation remained unchanged. Fig.~\ref{Fig-1StPairSizeComp} illustrates a comparison of the results for the two different crystal sizes and, as one can clearly see, the differences are extremely small, which is consistent with previous AIMD simulations of others \cite{Sugino1995}. Examining the the differences in the standard deviations for the two sets of the data, we conclude that all the digits reported in Table~\ref{Tab-Sigma} are fully converged with respect to the crystal size.

\section{Gibbs Measure: The Full Representation}
\label{Sec-QGibbsMeasure}

\begin{table}
\centering
\caption{The values of $\sigma_n$, $n=0,\ldots,4$, determining the Gibbs measure at the studied temperatures.}
\begin{tabular}{c c c c c c}
\hline\hline
 T (K)        & $\sigma_0$ (\AA) & $\sigma_1$ (\AA) & $\sigma_2$ (\AA) & $\sigma_3$ (\AA) & $\sigma_4$ (\AA)  \\
\hline
$300$       & 0.0593 & 0.0554 & 0.0376 & 0.0295 & 0.0238   \\
$600$       & 0.0874 & 0.0818 & 0.0552 & 0.0430 &  0.0346   \\
$900$       & 0.1123 & 0.1051 & 0.0710 & 0.0550 &  0.0444   \\
$1200$     & 0.1371 & 0.1293 & 0.0870 & 0.0669 &  0.0539   \\
$1500$     & 0.1630 & 0.1540 & 0.1034 & 0.0787 &  0.0633   \\
\hline
\end{tabular}
\label{Tab-Sigma}
\end{table}

In the previous section, we found that the Gibbs measure of the silicon crystal at various temperatures is extremely well characterized by a multivariate normal distribution of zero mean, which generically takes the form:
\begin{equation}
\PM({\rm d}\omega) = \rho(\omega) {\rm d}\omega, \quad \rho(\omega) = \tfrac{1}{\sqrt{{\rm Det}(2\pi \hat \Sigma)}}e^{-\frac{1}{2} \omega^T \hat \Sigma^{-1} \omega},
\end{equation}
where $\omega$ is seen here as a $1$-column matrix and $\hat \Sigma$ is the variance matrix. The entries of $\omega$ consists of the components $\omega_{\bm x}$, where $\bm x \in \Ll$, and each such component has three entries, $(\omega_{\bm x})_u$, with $u=x,y,z$. For example, the matrix product $\omega^T \hat \Sigma \omega$ takes the following explicit form:
\begin{equation}
\omega^T \hat \Sigma \omega = \sum_{\bm x_1,\bm x_2}\sum_{u_1,u_2} \Sigma_{\bm x_1,\bm x_2}^{u_1,u_2} (\omega_{\bm x_1})_{u_1} (\omega_{\bm x_2})_{u_2}.
\end{equation}
A fundamental property of normal distributions is that the variance matrix can be mapped out from the pair correlations:
\begin{equation}
\hat \Sigma_{\bm x_1,\bm x_2}^{u_1,u_2}=\lim_{T \rightarrow \infty}\tfrac{1}{T} \int_0^T {\rm d}t \, \omega_{\bm x_1}(t)_{u_1} \omega_{\bm x_2}(t)_{u_2} = \int_\Omega \PM({\rm d} \omega) \;  (\omega_{\bm x_1})_{u_1} (\omega_{\bm x_2})_{u_2}. 
\end{equation}
Going one step forward, we write the pair correlations in terms of the projected measures studied in section~\ref{Sec-HGibbsMeasure}:
\begin{equation}
\hat \Sigma_{\bm x,\bm x}^{u,v} = \int_{\Omega_0} \PM_0({\rm d} \omega_{\bm x}) \;  (\omega_{\bm x})_{u} (\omega_{\bm x})_{v}, \quad \bm x \in \Ll,
\end{equation}
and:
\begin{equation}
\hat \Sigma_{\bm x_1,\bm x_2}^{u_1,u_2} = \int_{\Omega_n} \PM_n({\rm d} \omega_{\bm x_1} \times {\rm d} \omega_{\bm x_2}) \;  (\omega_{\bm x_1})_{u_1} (\omega_{\bm x_2})_{u_2}, \quad (\bm x_1,\bm x_2) \in \Pp_n, 
\end{equation}
and now it becomes more clear that $\hat \Sigma$ can be mapped out explicitly from the analysis of section~\ref{Sec-HGibbsMeasure}. The result is:
\begin{equation}\label{Eq-FGibbsMeasure}
\omega^T \hat \Sigma \omega = \sigma^2_0\sum_{\bm x \in \Ll} \omega_{\bm x} \cdot \omega_{\bm x} + \sum_{n \geq 1} \sigma_n^2\sum_{(\bm x_1,\bm x_2)\in \Pp_n} \omega_{\bm x_1} \cdot \omega_{\bm x_2}, \quad \sigma_n = \sqrt{\beta_n^2 - \alpha_n^2}.
\end{equation}
A quick test of \eqref{Eq-FGibbsMeasure} consists of separating the block of the variance matrix $\hat \Sigma$ corresponding to a pair $(\bm x_1,\bm x_2) \in \Pp_n$ and compare the result with the ansatz \eqref{Eq-Ansatz2}. Recalling the relation $\sigma_0 = \sqrt{\tfrac{\alpha_n^2 + \beta_n^2}{2}}$, we find, after excluding all terms except those involving $\bm x_1$ and $\bm x_2$:
\begin{align}
\omega^T\, \hat \Sigma \, \omega \ \rightarrow \ & \sigma_0^2 \, (\omega_{\bm x_1} \cdot \omega_{\bm x_1} + \omega_{\bm x_2} \cdot \omega_{\bm x_2}) + \sigma_n^2 \,  \omega_{\bm x_1} \cdot \omega_{\bm x_2}
\\ \nonumber & = \tfrac{\alpha_n^2 + \beta_n^2}{2} (\omega_{\bm x_1} \cdot \omega_{\bm x_1} + \omega_{\bm x_2} \cdot \omega_{\bm x_2}) + (\beta_n^2 - \alpha_n^2)\omega_{\bm x_1} \cdot \omega_{\bm x_2}
\\ \nonumber & = \alpha_n^2 \, \tfrac{\omega_{\bm x_1} -  \omega_{\bm x_2}}{\sqrt{2}} \cdot \tfrac{\omega_{\bm x_1} -  \omega_{\bm x_2}}{\sqrt{2}} + \beta_n^2 \, \tfrac{\omega_{\bm x_1} +  \omega_{\bm x_2}}{\sqrt{2}} \cdot \tfrac{\omega_{\bm x_1} +  \omega_{\bm x_2}}{\sqrt{2}},
\end{align}
which is indeed fully consistent with the ansatz \eqref{Eq-Ansatz2}.

\begin{figure}
\center
  \includegraphics[width=0.9\textwidth]{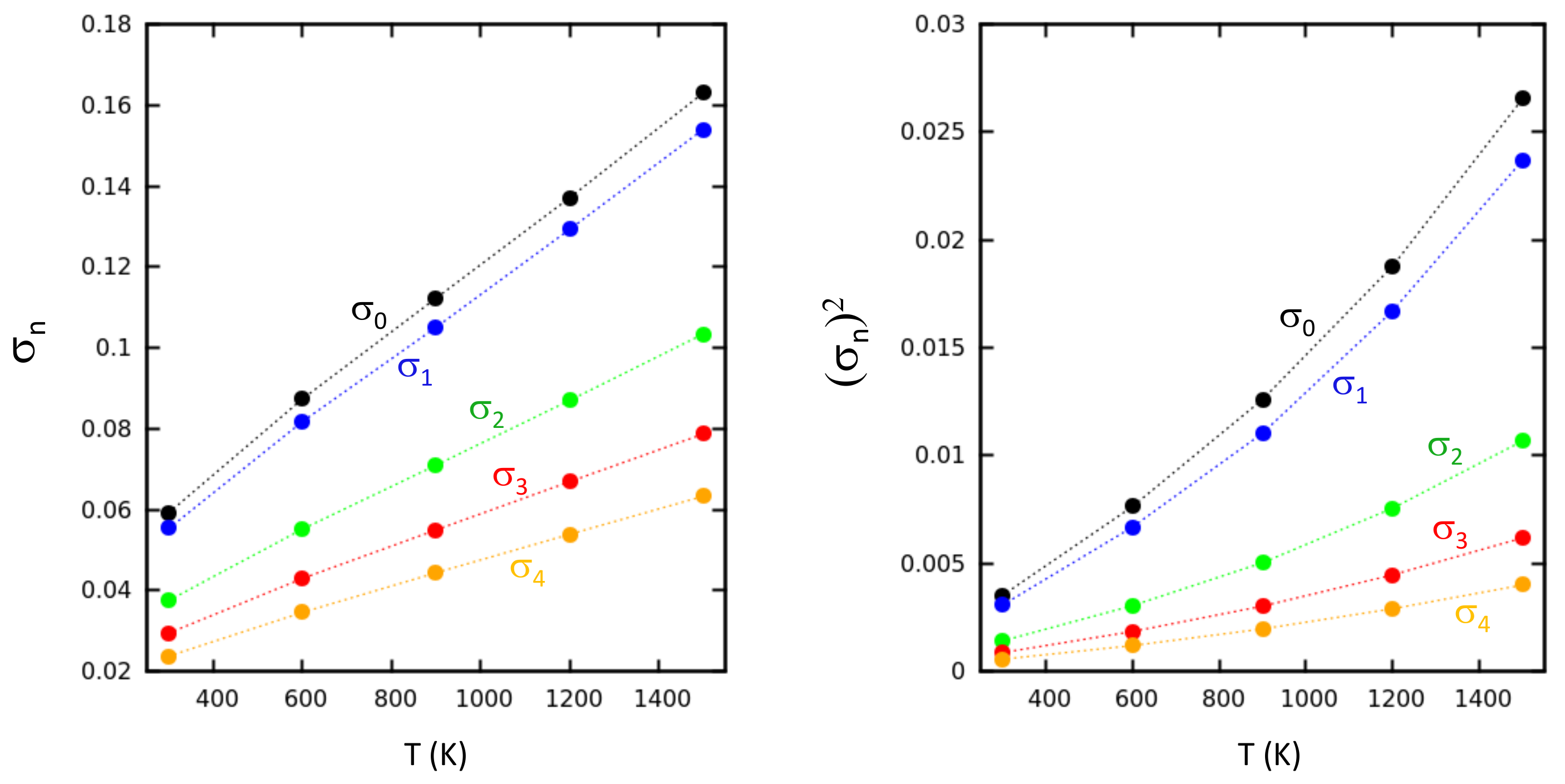}\\
  \caption{\small Dependence of $\sigma_n$ (left) and $(\sigma_n)^2$ (right), $n=0,\ldots,4$, on the temperature.}
 \label{Fig-SigmaVsTemp}
\end{figure}

\vspace{0.2cm}

At this point, we have reached the main conclusion of our work, namely, that the whole Gibbs measure can be fully encoded using just five parameter. These parameters are listed in Table~\ref{Tab-Sigma} for various temperatures. We mention that we applied the tests performed in section~\ref{Sec-HGibbsMeasure}, and found that, for all the temperatures reported in Table~\ref{Tab-Sigma}, they are passed with the same or even higher precision. We have excluded the temperature $1800 \ K$ from Table~\ref{Tab-Sigma} because of a poorer performance of the data under these tests. A plot of $\sigma_n$ and $\sigma_n^2$ as a function of temperature is shown in Fig.~\ref{Fig-SigmaVsTemp}, which demonstrates that the behavior of the coefficients is smooth with the temperature and, as such, the data can be interpolated and used for any temperature between 300 K and 1500 K.

\section{Conclusions}

The configuration space and the Gibbs measure of the atomic degrees of freedom of a crystal at finite temperature are complex objects which live in multi-dimensional spaces. In the present work, we realized that the projections of these two objects on various coordinate sub-spaces are numerically tractable and quantifiable in terms of a small number of parameters. By examining the various asymptotic behaviors, we reached the conclusion that the whole Gibbs measure can be re-constructed with extremely high precision from a finite number of such projections. 

\vspace{0.2cm} 

The {\it ab-initio} temporal orbits of similar or more complex crystals can be generated by the methods described in section~\ref{Sec-CPMD}. We now have an algorithmic method to analyze these orbits in order to quantify the configuration space and the Gibbs measure of the crystal. When the algorithm was applied to the silicon crystal, we found that the Gibbs measure is extremely well characterized by a multivariate normal distribution, whose covariance matrix can be fully specified by just five parameters. For more complex crystals, we expect this finding to hold true as well, though the number of parameters may be much larger. If these parameters can be tabulated, as done for the Silicon crystal in the present work, then accurate thermally disordered configurations of the crystals can be generated, {\it e.g.} using inexpensive classical Monte-Carlo simulations. The systems generated this way can be much larger than what can be achieved by direct {\it ab-initio} simulations. This is important because, while we found that the data for the atomic degrees of freedom of Si crystal is already converged for a 1000-atom super-cell,  the simulations of the electron transport will require a much larger crystal, and that can be indeed achieved now.

\vspace{0.2cm}

\section{Acknowledgements}

The authors would like to thank the Gauss Center for Supercomputing (GCS) for providing computing time through the John von Neumann Institute for Computing (NIC) on the GCS share of the supercomputer JUQUEEN at the J\"ulich Supercomputing Centre (JSC). Funding from the European Research Council (ERC) under the European Union's Horizon 2020 research and innovation programme (grant agreement No 716142) and from the Keck Foundation is kindly acknowledged.

\bibliographystyle{plain}

\end{document}